\documentclass[%
reprint,
amsmath,amssymb,
aps,
]{revtex4-1}

\usepackage{graphicx}
\usepackage{dcolumn}
\usepackage{bm}
\usepackage{hyperref}
\usepackage{color}
\usepackage{xcolor}
\usepackage{subfig}
\usepackage{multirow}
\usepackage{gensymb}

\def\f{\frac}

\def\be{\begin{equation}}
\def\ee{\end{equation}}
\def\bea{\begin{eqnarray}}
\def\eea{\end{eqnarray}}

\newcommand{\levicivita}{}
\def\levicivita#1#{\tensor#1{\epsilon}}

\begin{document}
	
	
	\title{Quasinormal modes and greybody factors of the novel four dimensional Gauss-Bonnet black holes in asymptotically de Sitter space time: Scalar, Electromagnetic and Dirac perturbations}
	
	\author{Saraswati Devi$^1$}
	\email{sdevi@iitg.ac.in}
	\author{Rittick Roy$^2$}
	\email{rittickrr@gmail.com}
	\author{Sayan Chakrabarti$^1$}%
	\email{sayan.chakrabarti@iitg.ac.in}
	\affiliation{$^1$Department of Physics, Indian Institute of Technology Guwahati, Guwahati 781039, India 
	}%
	\affiliation{$^2$Department of Physics, Indian Institute of Technology Bombay, Mumbai 400076, India 
	}%
	
	\date{\today}
	
	\begin{abstract}
		We find the low lying quasinormal mode frequencies of the recently proposed novel four dimensional Gauss-Bonnet de Sitter black holes for scalar, electromagnetic and Dirac field perturbations using the third order
		WKB approximation as well as Pad\'{e} approximation, as an improvement over WKB. We figure out the effect of the Gauss-Bonnet coupling $\alpha$ and the cosmological constant $\Lambda$ on the real and imaginary parts of the QNM frequencies. We also study the greybody factors and eikonal limits in the above background for all three different types of perturbations. 
	\end{abstract}
	
	\maketitle
	
	
	\section{Introduction}
	
	Black holes are one of the most intriguing objects in the theory of general relativity (GR). They are the simplest objects that one can come across in the study of GR, simply because they are parametrised by only three parameters: the mass, the charge and the spin. It is one of the reasons why black holes have attracted so much attention apart from the fact that they are mathematically beautiful as well as strange objects by their own merit at the same time. Among many other interesting areas of studies, quasinormal modes (QNMs) have gained attention for the last few decades in discussing the perturbations of black
	holes \cite{Kokkotas:1999bd,Nollert:1999ji,Berti:2009kk,Konoplya:2011qq}. It is well known that for a large family of black holes, the perturbation equations can be cast into a Schr\"{o}dinger like form. The QNMs come out as the solutions of the corresponding Schr\"{o}dinger like wave equation with complex frequencies with boundary conditions which are purely ingoing at the horizon and outgoing at spatial infinity. QN frequencies carry unique information about the black hole parameters and despite the classical origin, it was found that QNMs might provide a hint into the quantum nature of the black holes \cite{Hod:1998vk,Dreyer:2002vy,Maggiore:2007nq}. In addition, QNMs in anti de Sitter (AdS) space-time has been shown to appear naturally in the description of the dual conformal field theories living on the boundary (see \cite{Berti:2009kk} for a detailed list of references). QNMs of black holes have already been observed in the ground based experiments \cite{Abbott:2016blz,TheLIGOScientific:2016src} and they already present a plethora of information about black holes. However, research areas still remain open towards interpreting those results which requires the exploration of alternative theories of gravity \cite{Konoplya:2016pmh,Berti:2018vdi} towards understanding fundamental problems like singularity resolution or a quantum nature of gravity.
	
	QNMs of black holes have originally been studied in the context of Einstein's theory of general relativity (see \cite{Berti:2009kk} for a comprehensive list of references). 
	QNMs dominate the last stage of an extremely complicated yet intriguing process of the merger of binary compact objects (for example black hole (BH) - black hole or black hole - neutron star (NS) merger), whereas the first two stages consist of the inspiral of the two BHs or NSs and merger of the two BHs or NSs into a single one. It has been observed that the last stage of formation of the single BH or NS from the binary merger is dominated by the quasinormal ringing and this process corresponds to an extremely strong gravitational field which cannot be modeled using the help of post-Newtonian approximation. However, it is this last stage of the merger process which carries the necessary imprint of the characteristics of a particular theory of gravity \cite{Konoplya:2016pmh}. In fact black holes in a number of alternative theories of gravity may produce the same observational signature in the asymptotic regions, but can lead to qualitatively different features near the event horizon. Therefore, studying various alternative theories of gravity in the strong field region still remains an active and interesting area of research in the context of gravitational wave signatures of black holes. 
	
	One such alternative theory is the Einstein-Gauss-Bonnet (EGB) theory of gravity which consists of higher curvature corrections to the Einstein-Hilbert term in the gravitational action. Because of the reason described above, there had been a lot of interests in black holes arising from higher curvature corrections
	to Einstein-Hilbert action. On another front, lot of new developments in string theory \cite{Zwiebach:1985uq,Sen:2005iz,Moura:2006pz} had increased this particular theories importance on the gravity side as well. It is well known that low energy limits of string theories give rise to effective models of gravity in higher dimensions, which involve higher powers of the Riemann curvature tensor in
	the action in addition to the usual Einstein-Hilbert term \cite{Zwiebach:1985uq}. The Gauss-Bonnet combination $R^2-4R_{ab}R^{ab}+R_{abcd}R^{abcd}$ is of most interest among these higher powers of Riemann tensor and the theory also admits black hole solutions \cite{Boulware:1985wk,Wheeler:1985qd,Wiltshire:1988uq}. Not only in string generated gravity models, the Gauss-Bonnet black holes has also gained interest in the context of brane world models
	\cite{Meissner:2001xg} as well as in the context of possible production at the LHC \cite{Barrau:2003tk}.
	
	It is imperative to note that the Gauss-Bonnet action in $D$-space time dimensions, having the following form $\int d^Dx \sqrt{-g} \{R^2-4R_{ab}R^{ab}+R_{abcd}R^{abcd}\}$ gives non-trivial equations of motion only in $4+1$ dimensions or higher, while in $3+1$ dimensions the Gauss-Bonnet term reduces to a topological surface term (see \cite{Lovelock:1971yv,Lovelock:1972vz} for details). Lot of works has been done on quasinormal modes and stability of black holes arising out of EGB gravity in space time dimensions $D>4$\cite{Konoplya:2004xx,Abdalla:2005hu,Chakrabarti:2005cm,Grain:2005my,Chakrabarti:2006ei,Daghigh:2006xg,Konoplya:2008ix,Zhidenko:2008fp,Chakrabarti:2008xz,Konoplya:2010vz,Sadeghi:2011zza,Blazquez-Salcedo:2016enn,Graca:2016cbd,Maselli:2014fca,Konoplya:2017ymp,Gonzalez:2017gwa,Konoplya:2017zwo}.
	
	As mentioned, in four space time dimensions, the Gauss-Bonnet term does not contribute to the gravitational dynamics since it becomes a surface term. Although, the role played by the Gauss-Bonnet term in four dimensional gravity theories has been intriguing for a long time and few studies towards that direction could be found in \cite{Miskovic:2009bm, Araneda:2016iiy}. Very recently a non-trivial extension of EGB theory of gravity has been proposed by Glavan and Lin \cite{Glavan:2019inb} in four space time dimensions as $D\to 4$ limit of the higher dimensional Gauss-Bonnet theory. It has been shown that the EGB gravity theory can be reconstructed in a particular way where the Gauss-Bonnet coupling can be re-scaled as $\alpha/(D- 4)$, with $\alpha$ being the Gauss-Bonnet coupling. This theory in four space time dimension was soon termed as novel 4D EGB theory, which is defined as a $D\to 4$ limit at the level of equations of motion. It was shown that the $D\to 4$ singular limit of the Gauss-Bonnet term produces some non trivial contributions to the gravitational dynamics, but  preserves the number of graviton degrees of freedom. This novel EGB theory can be shown to be free from Ostrogradsky instability \cite{Glavan:2019inb} too. Moreover, it was shown that such a theory does not require coupling to any matter field, it bypasses all conditions imposed by Lovelock's theorem \cite{Lovelock:1971yv} and is also free from any singularity problem. The discovery of such a theory in $D=4$ dimensions, therefore, has generated tremendous interest in the area of higher curvature theories which has been reflected in the large volume of works being done in a short span of time \cite{Konoplya:2020bxa,Guo:2020zmf,Fernandes:2020rpa,Konoplya:2020qqh,Wei:2020ght,Hegde:2020xlv,Kumar:2020owy,Ghosh:2020vpc,Doneva:2020ped,Konoplya:2020ibi,Singh:2020xju,Ghosh:2020syx,Konoplya:2020juj,Zhang:2020qam,HosseiniMansoori:2020yfj,Roy:2020dyy,Singh:2020nwo,Churilova:2020aca,Mishra:2020gce,Islam:2020xmy,Konoplya:2020cbv,Liu:2020vkh,NaveenaKumara:2020rmi,Aragon:2020qdc,Kumar:2020xvu,Kumar:2020uyz,Malafarina:2020pvl,Yang:2020czk,Cuyubamba:2020moe,Mahapatra:2020rds,Rayimbaev:2020lmz,Liu:2020evp,Jusufi:2020yus,Zhang:2020sjh,Liu:2020yhu,EslamPanah:2020hoj,Heydari-Fard:2020sib}. It should be noted that since the proposal of this solution, there have been several studies which questions the existence of this particular solution \cite{Gurses:2020ofy, Ai:2020peo, Hennigar:2020lsl}. These studies argue that there is no consistent way of taking the  $D\rightarrow 4$ limit of a $D$ dimensional theory, as was suggested by Glavan and Lin. It was also shown in \cite{Shu:2020cjw}, by considering a semi-classical quantum tunnelling of the vacuum of the solution, that the vacuum decay rate would exhibit diverging behaviour or complex behaviour, leading to an unstable or non-physical vacuum of the 4D Einstein Gauss-Bonnet solution. Since then, several regularization schemes have been proposed in order to overcome these shortcomings \cite{Lu:2020iav, Kobayashi:2020wqy, Fernandes:2020nbq}. Note that in this work, we would proceed by considering the solution as proposed by Glavan and Lin. Our aim in this work is to study the quasinormal modes of black holes in four dimensional novel EGB gravity in asymptotically de Sitter spaces. While there were many works on constructing different black hole solutions, such as static spherically symmetric black holes \cite{Glavan:2019inb}, black holes in AdS spaces \cite{Fernandes:2020rpa}, rotating black holes and their shadows \cite{Wei:2020ght,Kumar:2020owy}, generalised four dimensional black holes in Einstein-Lovelock gravity \cite{Konoplya:2020qqh}, radiating Vaidya like black holes \cite{Ghosh:2020syx}, regular black holes \cite{Kumar:2020uyz,Kumar:2020xvu}; not much effort has gone into figuring out the QNMs of spherically symmetric black holes with the exceptions of \cite{Konoplya:2020bxa,Churilova:2020aca,Mishra:2020gce,Cuyubamba:2020moe}. Our aim is to fill up this gap in the literature by studying the quasinormal modes of spherially symmetric black hole in novel four dimensional EGB gravity in asymptotically de Sitter space time.  
	
	The plan of this paper is as follows: in the next section we will briefly discuss about the four dimensional EGB gravity and the black hole solutions in them with a particular focus on the asymptotically de Sitter branch. In section III, we will describe the scalar, electromagnetic and Dirac perturbations of the black hole metric and will briefly describe the methodology adopted to evaluate the QN frequencies, section IV presents the results of our calculations. We give a brief discussion on the eikonal limit, Lyapunov exponents and unstable circular null geodesics following \cite{Cardoso:2008bp} in section V. A very brief discussion on the greybody factors is given in section VI. Finally we conclude the paper with a discussion on our results and future outlook.

	\section{Novel four dimensional Einstein-Gauss-Bonnet gravity}
	
	In their recent work, Glavan and Lin \cite{Glavan:2019inb} had shown by constructing a model of the novel four dimensional Einstein-Gauss-Bonnet gravity that the four important criteria, dictated by Lovelock's theorem \cite{Lovelock:1971yv,Lovelock:1972vz} for Einstein's general relativity with the cosmological constant to be an unique theory of gravity (viz. existence of 3+1 dimensional space time, general coordinate invariance, metricity and existence of second order equations of motion) can be overridden and the model can exhibit modified dynamics. To understand it in a better way, let us recall that the action for a general $D$-dimensional EGB gravity theory can be written as
	\bea
	S_{EGB}[g_{ab}]=S_{EH}[g_{ab}]+S_{GB}[g_{ab}]\label{action},
	\eea
	where the Einstein-Hilbert action is 
	\bea
	S_{EH}[g_{ab}]=\f{1}{16\pi G_N}\int d^Dx\sqrt{-g}\left[R-2\Lambda\right].\label{ehaction}
	\eea
	In writing Eq. (\ref{ehaction}) and in the rest of the paper, we have chosen $G_N$, the $D$-dimensional Newton's constant to be unity, $R$ is the Ricci scalar and $\Lambda$ is the bare cosmological constant. In fact both $G_N=1/8\pi M_{\rm{pl}}^2$ and $\Lambda$ are the parameters of the theory, where $M_{\rm{pl}}$ is the Planck mass characterising the strength of the gravitational interaction. It is well known that Einstein's General theory of Relativity is a perturbatively non-renormalizable theory, and it can be made sensible by adding higher curvature corrections to the Einstein-Hilbert action in strong gravity regimes. Among different choices of higher curvature terms, the Lovelock corrections play a crucial role in the sense that the field equations contain terms only up to the second derivative of the metric and secondly, the gravitational dynamics remains free of the Ostrogradsky instabilities. Of particular interest is the third order Lovelock correction, known as the Gauss-Bonnet term and the action looks like
	\bea
	S_{GB}[g_{ab}]=\f{1}{16\pi}\int d^Dx \sqrt{-g} \alpha \mathcal{G}, \label{gbaction}
	\eea
	where, $\alpha$ is the Gauss-Bonnet coupling constant and $\mathcal{G}$ is the Gauss-Bonnet term having the form $\mathcal{G}=R_{abcd}R^{abcd}-4R_{ab}R^{ab}+R^2$ in which $R_{abcd}$ is the Riemann curvature tensor, $R_{ab}$ is the Ricci tensor and $R$ is the Ricci scalar. Incorporating such terms in the Einstein-Hilbert action had already generated many interesting scenarios a few of which was mentioned in the introduction of this paper. It is to be noted that the Gauss-Bonnet term in four space time dimensions turns out to be a total divergence, hence it does not contribute to any gravitational dynamics. However, by re-scaling the Gauss-Bonnet coupling constant in $\mathcal{G}$ as $\alpha\to \alpha/(D - 4)$
	and then taking the limit $D\to 4$, one can obtain the novel four dimensional EGB gravity theory \cite{Glavan:2019inb}. Therefore, following Eqn. (\ref{action}), the action for novel four dimensional EGB gravity with the scaled coupling constant $\alpha/(D - 4)$ can be written as
	\bea
	S_{EGB}[g_{ab}]&&=\f{1}{16\pi}\int d^Dx\sqrt{-g}\Big[R-2\Lambda +\f{\alpha}{D-4}\times \nonumber\\ 
	&&(R_{abcd}R^{abcd}-4R_{ab}R^{ab}+R^2)\Big]+S_{\rm{matter}},
	\eea
	where, $S_{\rm{matter}}$ represents action corresponding to any matter fields in the theory. One can vary the action with respect to the metric and setting the variation to be equal to zero: $\delta S_{EGB}/\delta g^{ab}=0$, leads to the equations of motion
	\bea
	8\pi T_{ab}=G_{ab}^{(\Lambda)}+G_{ab}^{(EH)}+G_{ab}^{(LL)},\label{eom}
	\eea
	where, the tensors in the RHS of Eqn. (\ref{eom}) respectively are $G_{ab}^{(\Lambda)}=\Lambda g_{ab}$, the Einstein tensor $G_{ab}^{(EH)}=R_{ab}-\f{1}{2}R g_{ab}$  and the Lanczos-Lovelock tensor 
	\bea
	&&G_{ab}^{(LL)}= -\f{\alpha}{D-4}\Big[\f{1}{2}g_{ab}(R_{mnpq}R^{mnpq}-4R_{mn}R^{mn}+R^2)\nonumber\\
	&&-2RR_{ab}-4 R_{am}R^m_{~b}+4R_{ambn}R^{mn}-2R_{amnp}R^{~mnp}_{b}\Big].
	\eea
	The four dimensional novel EGB theory, at the level of equations of motion, can be obtained as a limit $D\to 4$ \cite{Glavan:2019inb}, circumventing the Lovelock's theorem. Such a theory admits black hole solutions (it admits both de Sitter as well as anti de Sitter branches, see \cite{Glavan:2019inb,Fernandes:2020rpa,Ghosh:2020syx} for details):
	\bea
	ds^2=-f(r)dt^2+f^{-1}(r)dr^2+r^2(d\theta^2+\sin^2\theta d\phi^2)\nonumber,
	\eea
	with
	\bea
	f(r)=1+\f{r^2}{32\pi\alpha}\left[1-\sqrt{1+\f{128\pi\alpha M}{r^3}+\f{64\pi\alpha\Lambda}{3}}\right]. \label{metric}
	\eea
	In the above $M$ is related to the black hole mass. In the limit $\alpha\to 0$, the above solution
	reduces to the  Schwarzschild de Sitter solution and as  $r\to\infty$, $f(r)$ reduces to the asymptotically de Sitter space time with positive cosmological constant. The Gauss-Bonnet coupling constant $\alpha$ can in principle be either positive or negative. In fact, it can be shown that in appropriate parameter region, the solution has two horizons: the event horizon $r_H$ and the cosmological horizon $r_c$. However, in the $\alpha <0$ regime, the metric function does not remain real for small values of the radial coordinate. However, we are not interested in very small values of $r$, rather our interest lies in the region  $r_H < r< r_c$, therefore, we  can in principle allow $\alpha$ to take negative values.

	\section{Scalar, electromagnetic and Dirac perturbations}
	We study the quasinormal modes of the metric given by Eq. (7) for scalar, electromagnetic and Dirac perturbations. Here we will take a rescaling of the Gauss-Bonnet coupling constant $32\pi\alpha\rightarrow \alpha$, and use this as the new Gauss-Bonnet coupling constant for convenience. So the metric simply reduces to
	\begin{align}
		f(r)= 1+\frac{r^2}{\alpha}\bigg[1-\sqrt{1+\frac{4\alpha M}{r^3}+{\frac{2\alpha{\Lambda}}{3}}}\bigg].
	\end{align}
	The Klein-Gordon equation for a massless scalar field in a black hole background takes the form
	\begin{align}
		\frac{1}{\sqrt{-g}}\partial_{a}(\sqrt{-g}g^{ab}\partial_{b}\Phi)=0, 
	\end{align}
	whereas, the electromagnetic field in curved spacetime follows the equation
	\begin{align}
		\frac{1}{\sqrt{-g}}\partial_{a}(\sqrt{-g} F_{bc}g^{bd}g^{ca})=0,
	\end{align}
	where $F_{bc}=\partial_{b}A_{c}-\partial_{c}A_{b}$ and $A_{a}$ is the four vector potential. After separation of variables, the radial parts of the above equations take the form
	\begin{equation}
	\frac{d^{2}\Psi_{s}}{dr_{*}^{2}}+(\omega^{2}-V_{s}(r))\Psi_{s}=0
	\label{eq:master}
	\end{equation}
	where, $s =$ \textit{``scalar"} refers to scalar field and $s =$ \textit{``em"} refers to electromagnetic field and $r_{*}$ is the tortoise coordinate defined as
	\begin{align}
		dr_{*}=\frac{dr}{f(r)}
	\end{align}
	The effective potentials for the scalar and electromagnetic cases are respectively given by
	\begin{align}
		\label{Vs}
		V_{{\rm{scalar}}}(r)=f(r)\bigg(\frac{l(l+1)}{r^{2}}+\frac{1}{r}\frac{df(r)}{dr}\bigg), 
	\end{align}
	and
	\begin{align}
		\label{Vem}
		V_{{\rm{em}}}(r)=f(r)\bigg(\frac{l(l+1)}{r^{2}}\bigg).
	\end{align}
	For a Dirac field on the other hand, the covariant Dirac equation has the form \cite{Brill:1957fx}
	\begin{equation}\label{covdirac}
	\gamma^{\alpha} \left( \frac{\partial}{\partial x^{\alpha}} - \omega_{\alpha} \right) \Psi=0,
	\end{equation}
	where $\gamma^{\alpha}$ are the Dirac gamma matrices and $\omega_{\alpha}$ are the spin connections. After applying the method of separation of variables, the radial part of the above equation can be cast in the Schr\"{o}dinger like form again, however with two different potentials corresponding to two different chiralities
	\begin{equation}  \label{wave-equation}
	\dfrac{d^2 \Psi_\pm}{dr_*^2}+(\omega^2-V_\pm^{{\rm{dirac}}}(r))\Psi_\pm=0,
	\end{equation}
	where, the effective potentials are of the form:
	\begin{equation}
	\label{Vd}
	V_{\pm}^{{\rm{dirac}}}(r) = \frac{(l+1)}{r}f(r) \left(\frac{(l+1)}{r}\mp\frac{\sqrt{f(r)}}{r}\pm \f{d\sqrt{f(r)}}{dr}\right).
	\end{equation}
	Note that the potentials $V_+^{dirac}(r)$ and $V_-^{{\rm{dirac}}}(r)$ can be transformed among each other implying that the quasinormal modes obtained from these two seemingly different potentials are isospectral.  Therefore one can use either of the two  $V_{\pm}^{{\rm{dirac}}}(r)$ for calculation of quasinormal modes.
	\begin{figure*}
		\centering
		\subfloat[$\alpha=0.2$ (red) and $\alpha=-0.2$ (blue) ]{{\includegraphics[width=8cm]{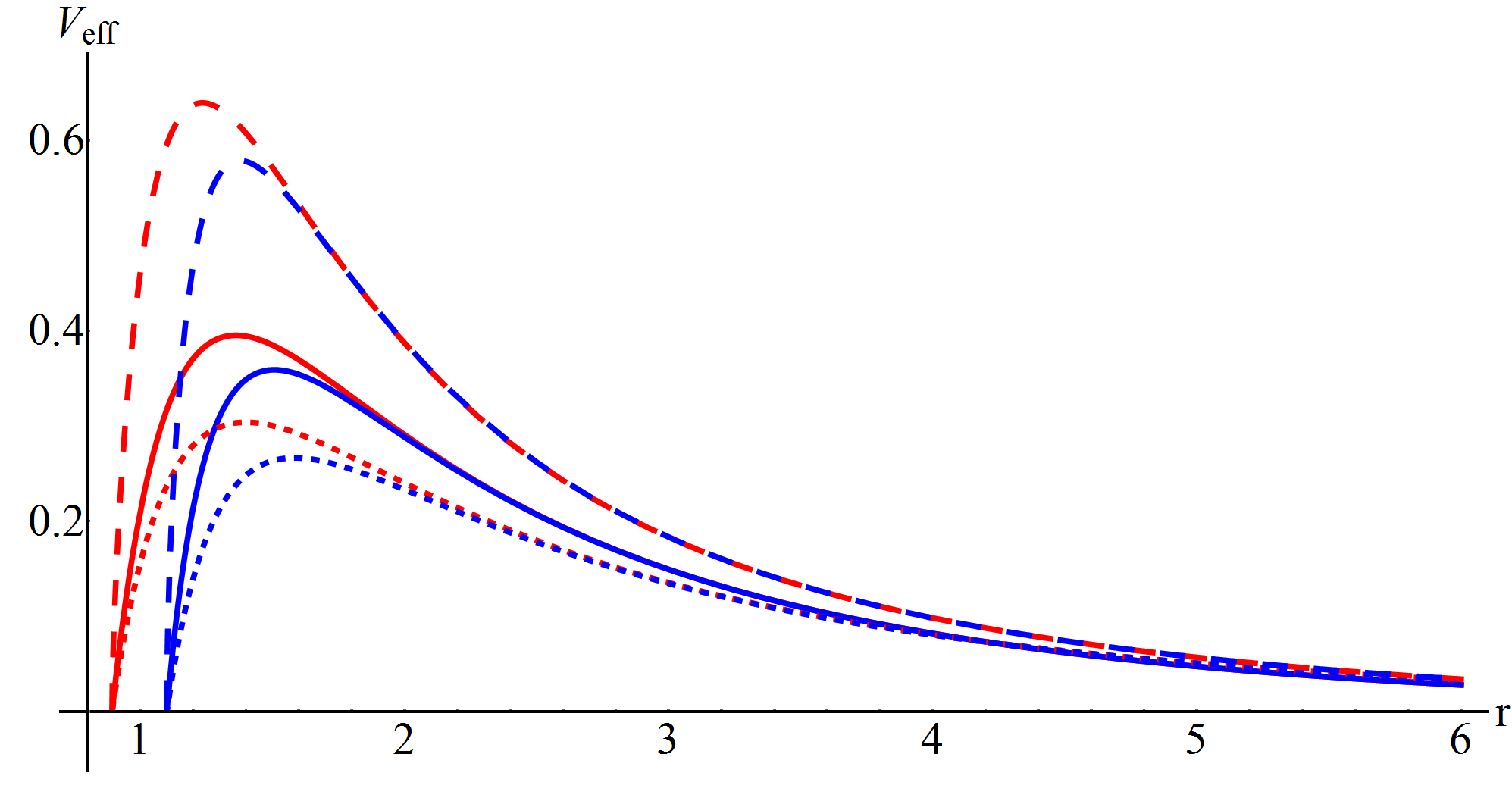} }}%
		\qquad
		\subfloat[$\Lambda=0.02$ (red) and $\Lambda=0.08$ (blue)]{{\includegraphics[width=8cm]{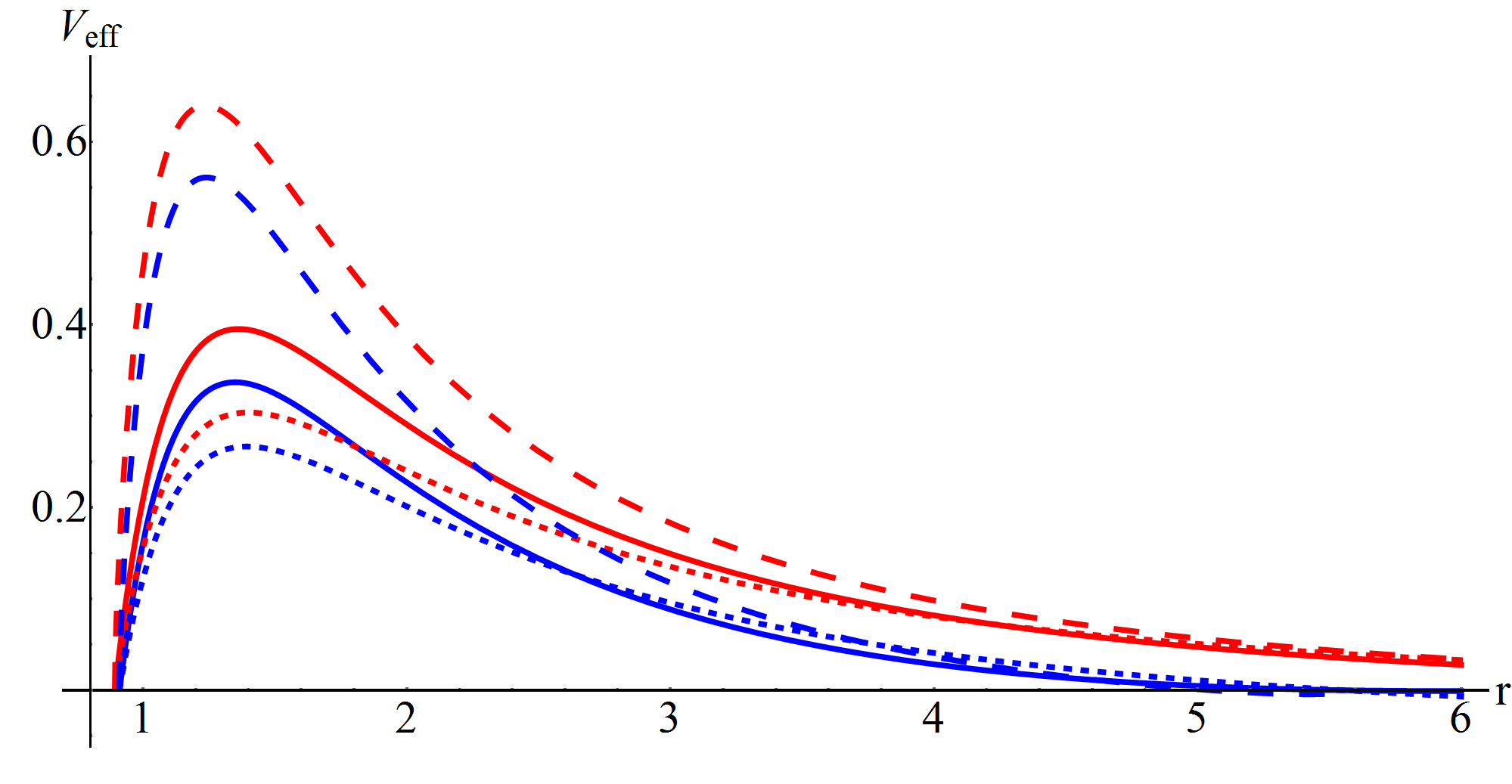} }}%
		\caption{The figure plots the effective potential $V_{eff}$ with the radial coordinate $r$ for the scalar (solid), electromagnetic (dotted) and the Dirac (dashed) perturbation.}%
		\label{fig:Veff}%
	\end{figure*}
	It is worth mentioning here that the stability of the scalar and electromagnetic perturbations in a general black hole background can be confirmed from the positive definiteness of the effective potential. However, it was shown very recently that the situation with the Dirac field is a little bit different, particularly if one considers higher curvature corrected black holes as well as study them in asymptotically de Sitter space times. Firstly, it was shown that even if the effective potential for one of the chiralities consists of a negative gap, the Dirac field perturbation can keep the black hole stable \cite{Zinhailo:2019rwd}. However, the positive definiteness of the potential of any one of the potentials for any one of the chiralities does not help in asymptotically de Sitter black hole backgrounds because the potential for both chiralities in general may have negative gaps \cite{Konoplya:2020zso}. Keeping these features in mind, we plan to study the quasinormal modes of the novel Gauss-Bonnet de Sitter black hole in four space time dimensions. We plot the effective potential of all three kinds of perturbations in Fig. (\ref{fig:Veff}). The quasinormal modes are the solutions of the master wave equation given by Eq. (\ref{eq:master}) satisfying the conditions of purely outgoing waves at infinity and pure ingoing waves at the event horizon. In the next section we will look into approximation routines to compute the quasinormal frequencies for the above three types of perturbations.
	
	\subsection{Methodology used: WKB approximation and Pad\'{e} approximation}
	It is well known that the analytic computation of quasinormal modes is almost impossible in most of the cases except a few background like BTZ black hole. Therefore, in order to numerically obtain the quasinomal frequencies, we have employed the 3rd order WKB approximation along with the improvements figured out using 3rd order Pad\'{e} approximation. It is already very well known that based on the semi classical arguments, Schutz and Will \cite{Schutz:1985zz} had suggested the WKB technique, which was later modified in \cite{Iyer:1986np}, by matching the exterior WKB solutions across the two turning points, which can be done only when the two classical turning points are close enough (see \cite{bender} for more details). The potential in the interior region was then expanded using the Taylor series expansion upto sixth order. The asymptotic approximation to the interior solution is used to match the 3rd order WKB. The formula for quasinormal frequencies using third order WKB approach is given by \cite{Iyer:1986np}
	\be
	\omega^2=[V_0+(-2V_0^{\prime\prime})^{1/2}\tilde\Lambda(n)]-i(n+\frac{1}{2})(-2V_0^{\prime\prime})^{1/2}[1+\tilde\Omega(n)].\label{freq}
	\ee
	where, $\tilde\Lambda=\Lambda/i$ and
	$\tilde\Omega=\Omega/(n+\frac{1}{2})$ and $\Lambda$ and $\Omega$ are
	given by
	\bea
	\Lambda(n)&=&\frac{i}{(-2V^{\prime\prime}_0)^{1/2}}\Big[\frac{1}{8}\left(\frac{V^{(4)}_0}{V^{\prime\prime}_0}\right)
	\left(\frac{1}{4}+\nu^2\right)\nonumber\\
	&&-\frac{1}{288}\left(\frac{V^{(3)}_0}{V^{\prime\prime}_0}\right)^2
	(7+60\nu^2)\Big]
	\eea
	\bea
	\Omega(n)&=&\frac{(n+\frac{1}{2})}{(-2V^{\prime\prime}_0)^{1/2}}\bigg [\frac{5}{6912}
	\left(\frac{V^{(3)}_0}{V^{\prime\prime}_0}\right)^4
	(77+188\nu^2)\nonumber \\
	&&-\frac{1}{384}\left(\frac{V^{(3)^2}_0V^{(4)}_0}{V^{\prime\prime^3}_0}\right)
	(51+100\nu^2)+\frac{1}{2304}\times\nonumber\\
	&&\left(\frac{V^{(4)}_0}{V^{\prime\prime}_0}\right)^2(67+68\nu^2)+\frac{1}{288}
	\left(\frac{V^{(3)}_0V^{(5)}_0}{V^{\prime\prime^2}_0}\right)\times\nonumber\\&& (19+28\nu^2)-\frac{1}{288}
	\left(\frac{V^{(6)}_0}{V^{\prime\prime}_0}\right)(5+4\nu^2)\bigg ],\label{wl}
	\eea
	where, $V^{(n)}_0$ is the $n$-th derivative of the effective potential with respect to the 
	tortoise coordinate calculated at the maximum of the potential $r_0$, $V_{0}$ is the height of the potential maximum and  $\nu=n+1/2$, where $n$ is a positive integer.
	\begin{figure*}%
		\centering
		\subfloat[Scalar perturbation]{{\includegraphics[width=5.43cm]{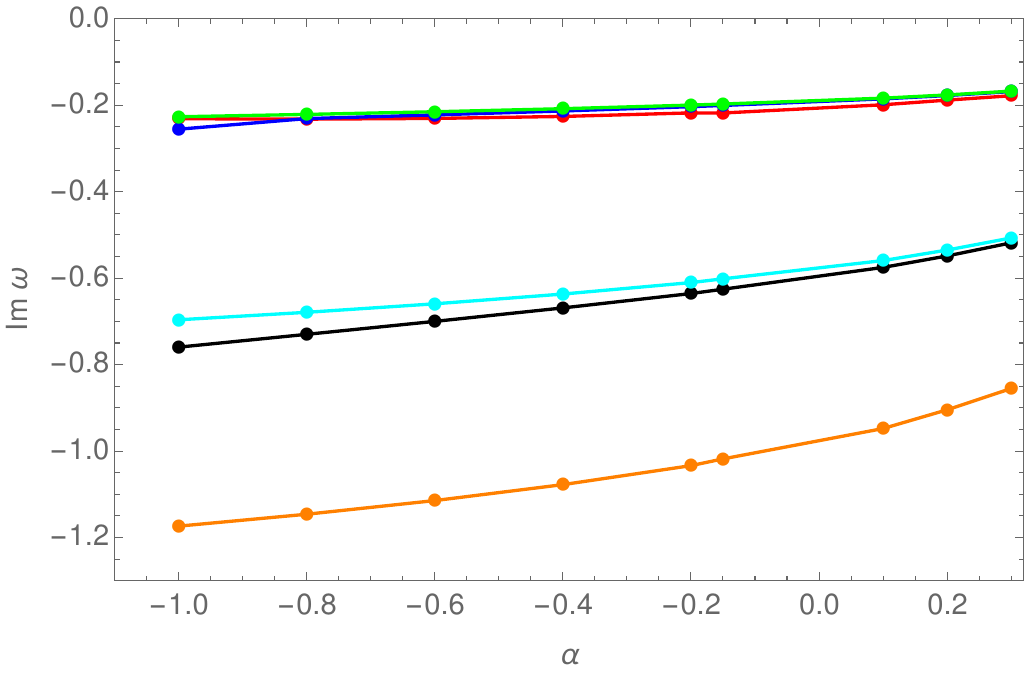} }}%
		\qquad
		\subfloat[Electromagnetic perturbation]{{\includegraphics[width=5.43cm]{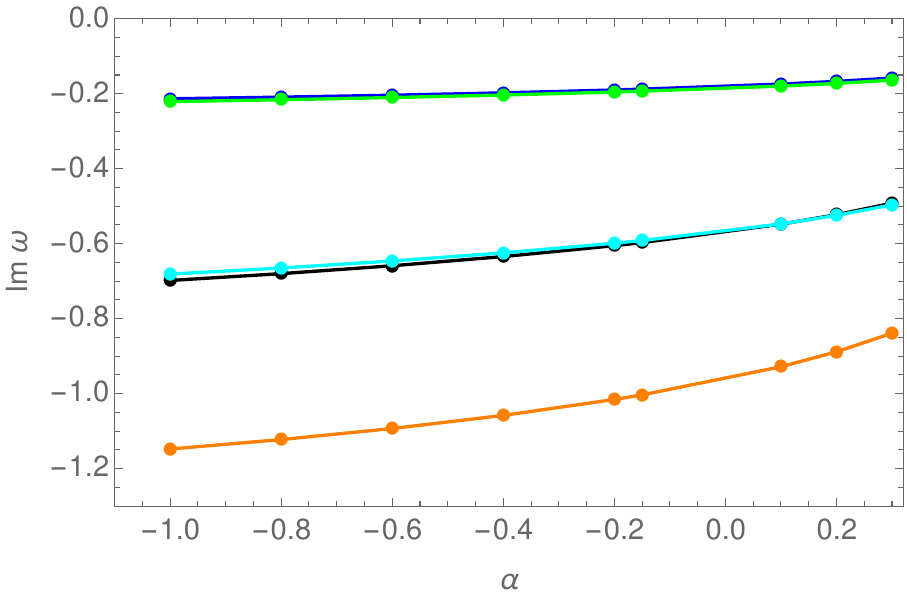} }}%
		\qquad
		\subfloat[Dirac perturbation]{{\includegraphics[width=5.43cm]{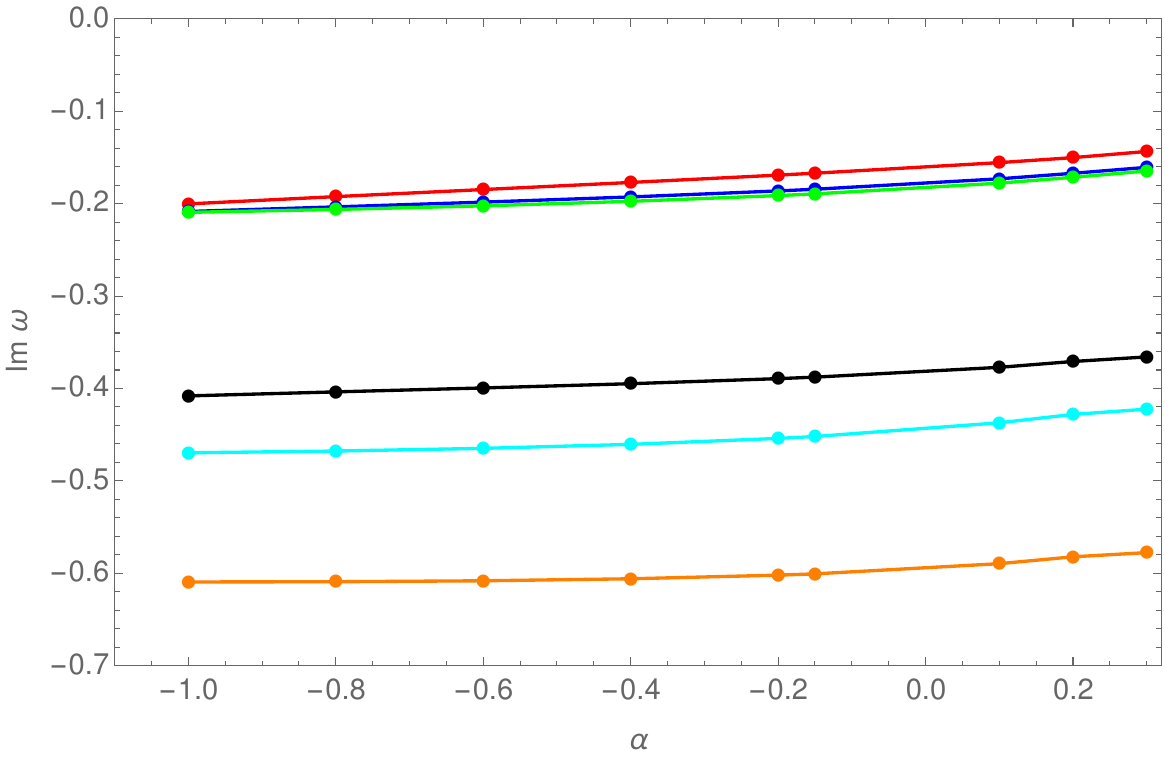} }}%
		\qquad
		\subfloat[Scalar perturbation]{{\includegraphics[width=5.43cm]{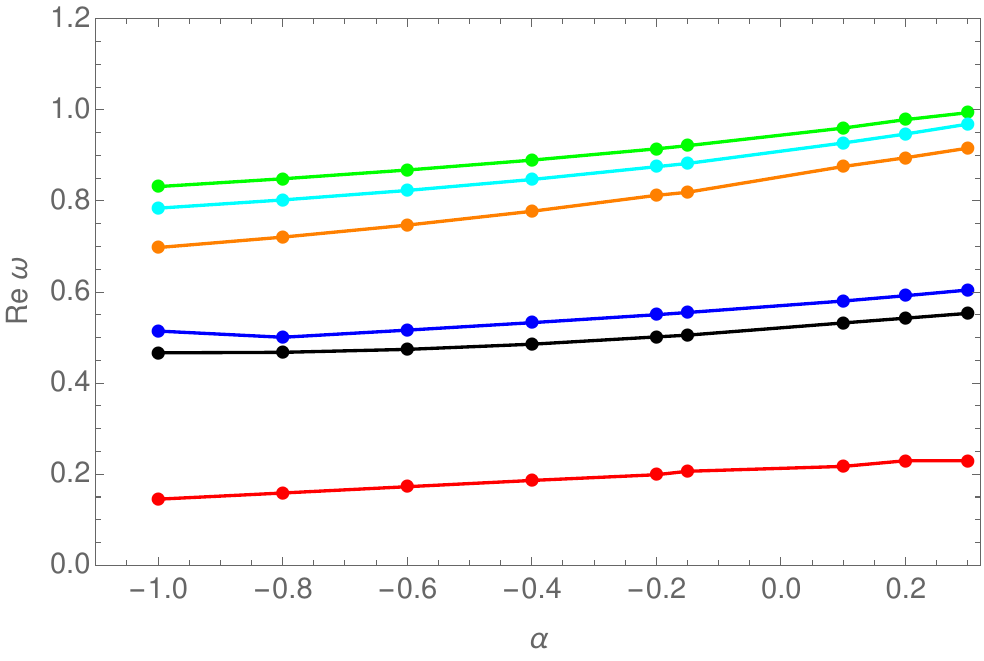} }}%
		\qquad
		\subfloat[Electromagnetic perturbation]{{\includegraphics[width=5.43cm]{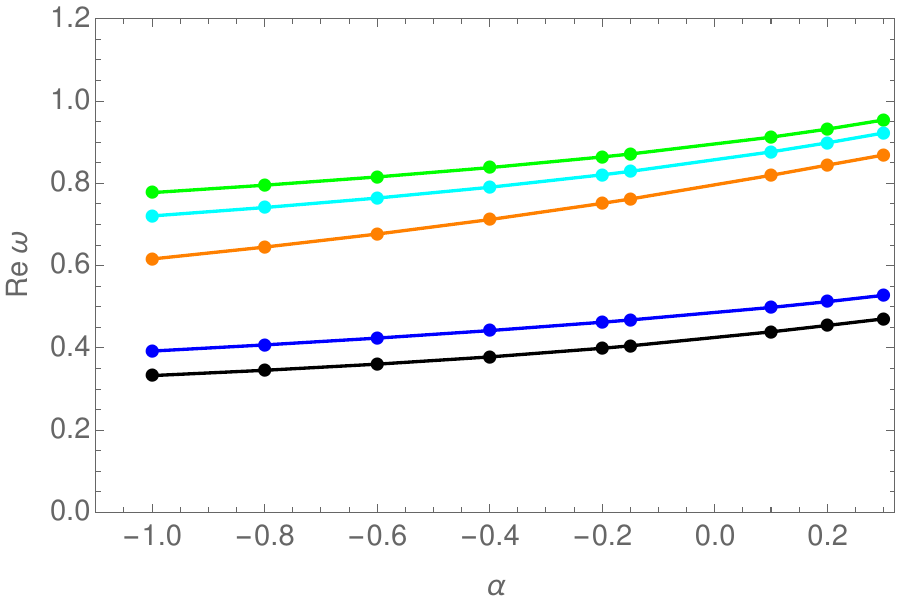} }}%
		\qquad
		\subfloat[Dirac perturbation]{{\includegraphics[width=5.43cm]{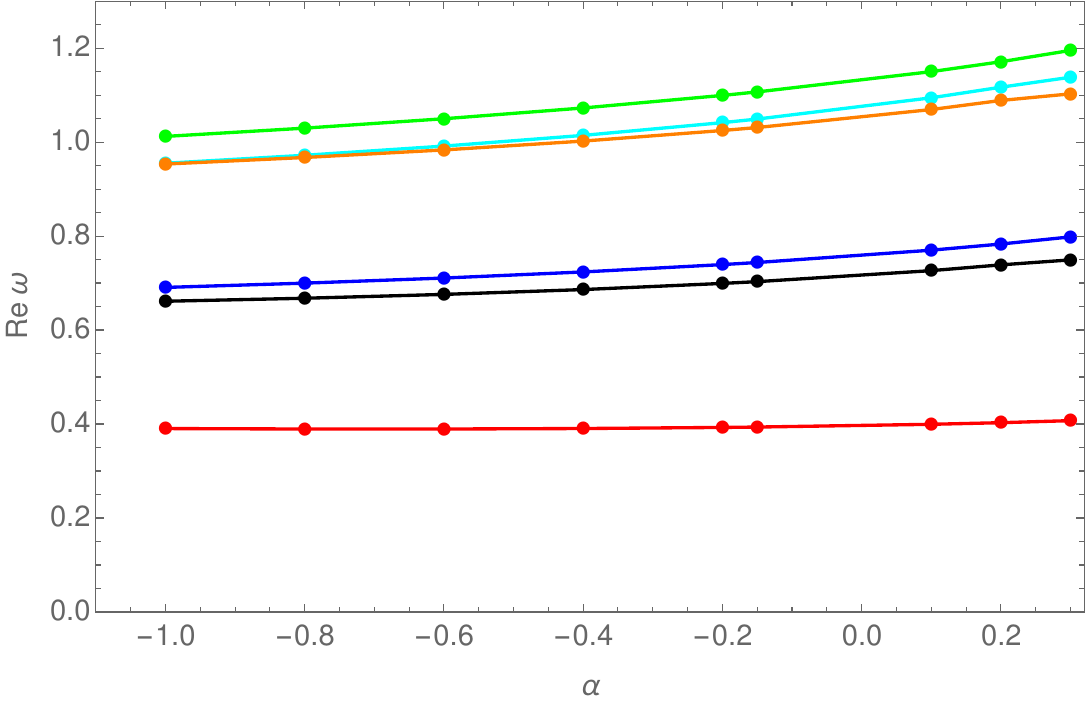} }}%
		\qquad
		\subfloat[Scalar perturbation]{{\includegraphics[width=5.43cm]{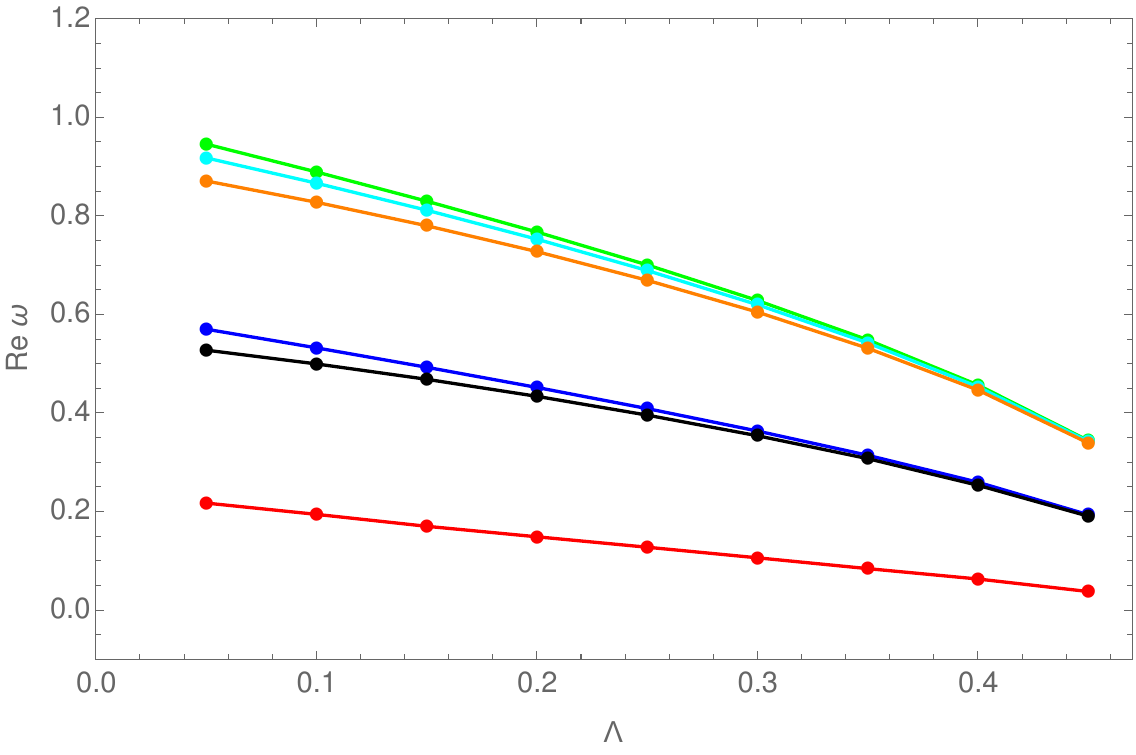} }}%
		\qquad
		\subfloat[Electromagnetic perturbation]{{\includegraphics[width=5.43cm]{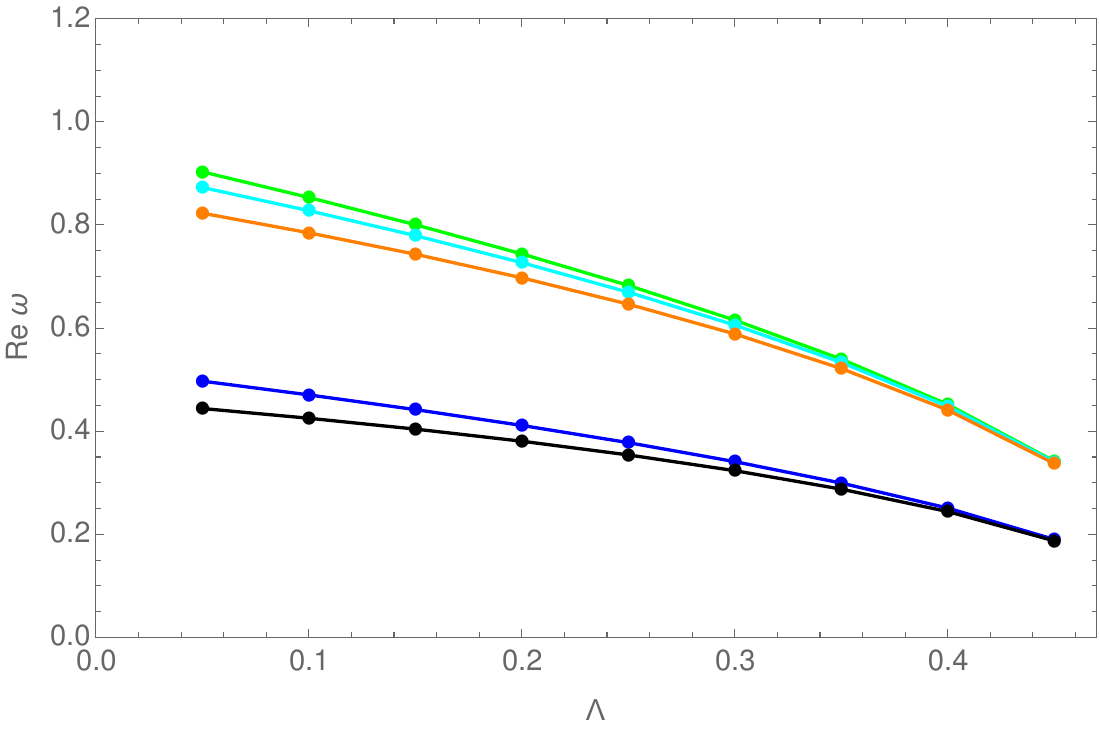} }}%
		\qquad
		\subfloat[Dirac perturbation]{{\includegraphics[width=5.43cm]{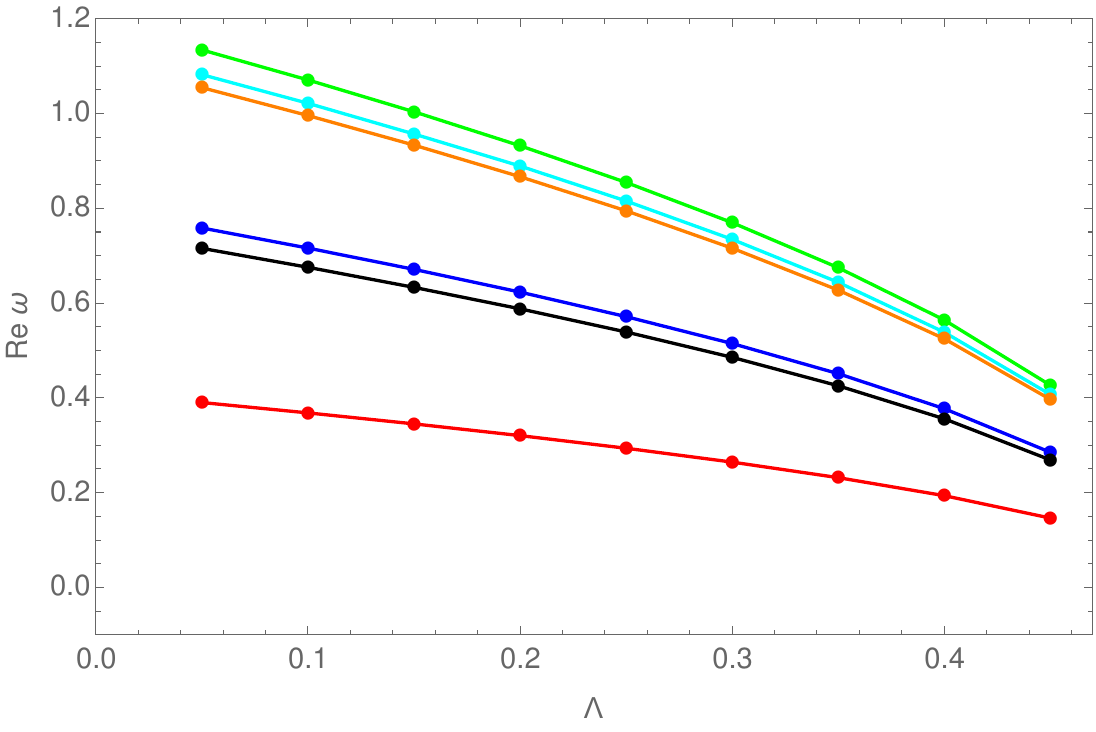} }}%
		\qquad
		\subfloat[Scalar perturbation]{{\includegraphics[width=5.43cm]{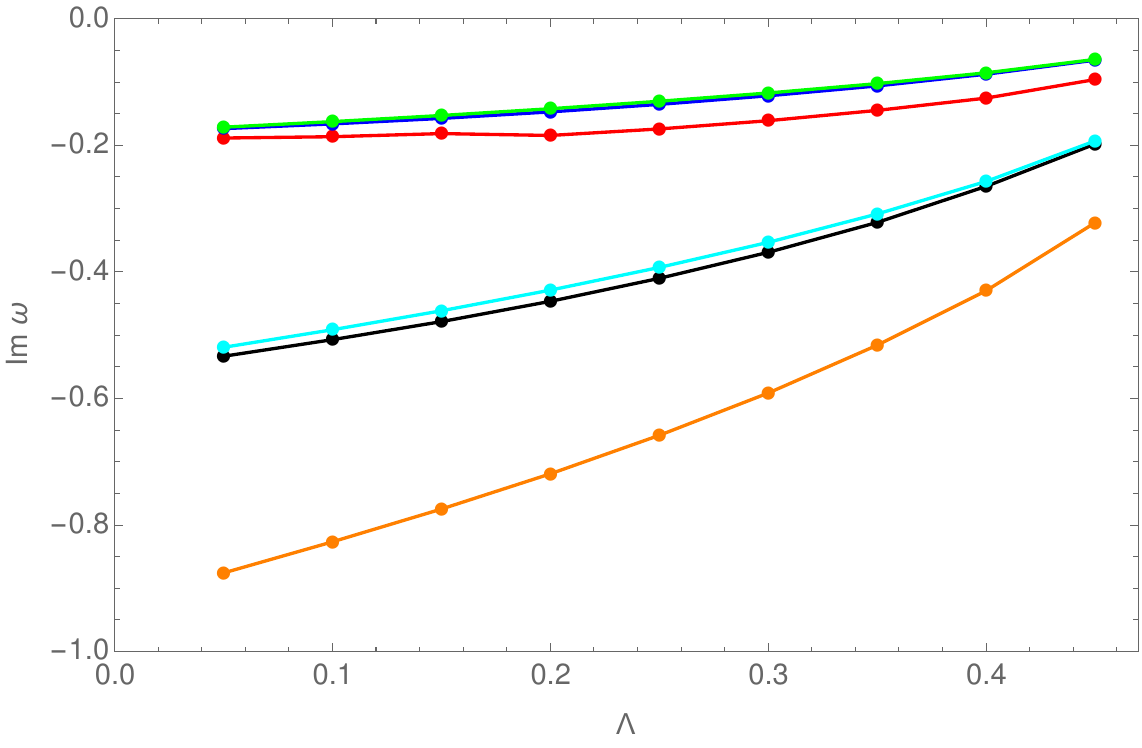} }}%
		\qquad
		\subfloat[Electromagnetic perturbation]{{\includegraphics[width=5.43cm]{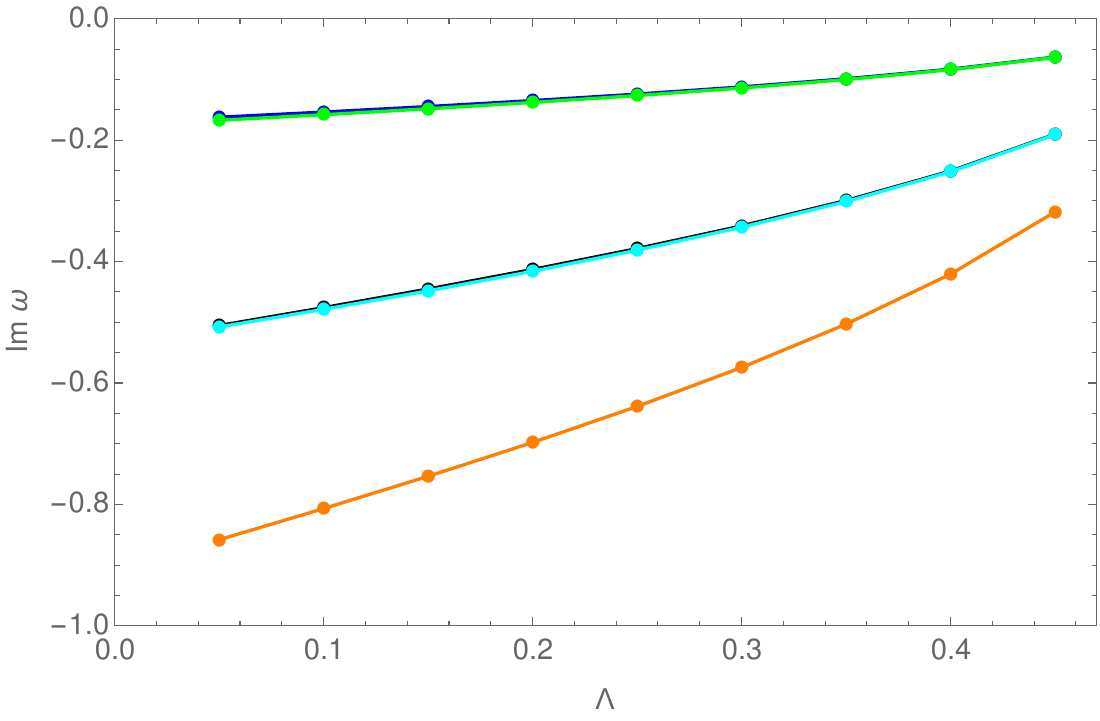}  }}%
		\qquad
		\subfloat[Dirac perturbation]{{\includegraphics[width=5.43cm]{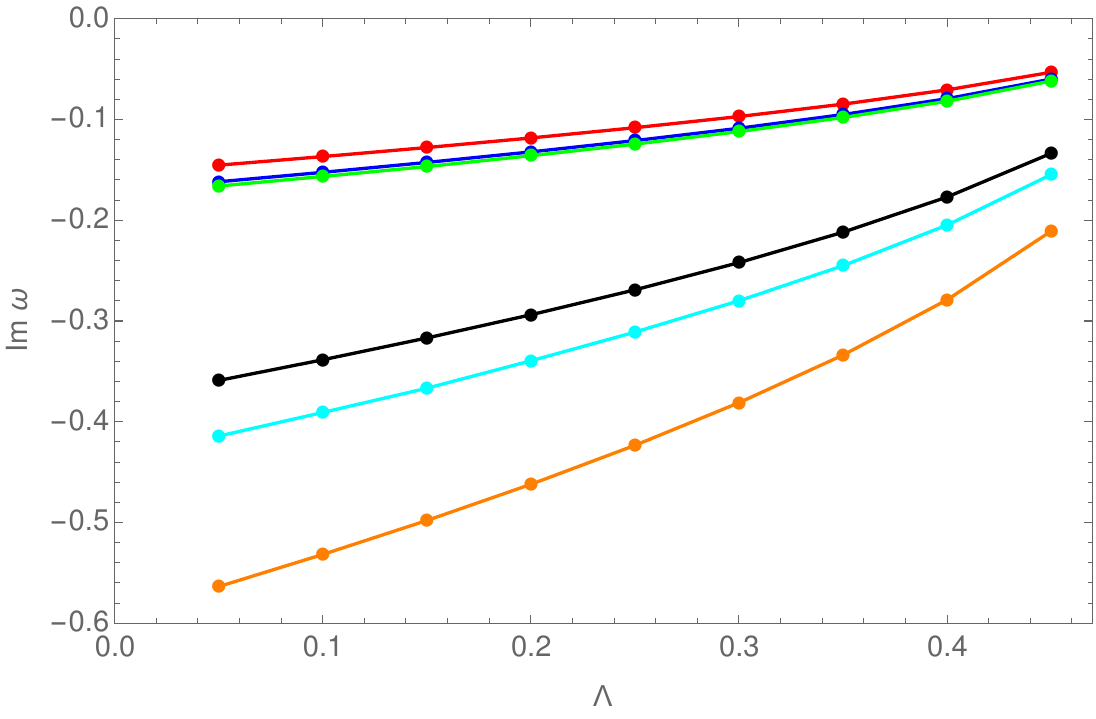} }}%
		\caption{The figure plots the real and imaginary parts of the quasinormal frequencies vs the different parameters $(\alpha,\Lambda)$ of the four dimensional Gauss-Bonnet de Sitter spacetime, for different types of perturbations, for a fixed mass of $M=0.5$. The different colors denote the different modes for different values of $(l,n)$: red $(0,0)$; blue $(1,0)$; black $(1,1)$; green $(2,0)$; cyan $(2,1)$ and orange $(2,2)$}%
		\label{fig:qnf}%
	\end{figure*}
	As a matter of fact, it should be pointed out here that the accuracy of the WKB method depends crucially on the multipole number $l$ and the overtone number $n$. It has been shown in \cite{Cardoso:2004fi} that the WKB approach works extremely well for situations where the multipole number is larger compared to the overtone: $l>n$. It is such a good approximation that the results from numerical integration of the wave equation and the WKB results are in good agreement, but the WKB approach does not yield satisfactory results if $l=n$ and is not applicable for $l<n$. On the other hand, the results are progressively better with increasing $l$ values. In order to increase the accuracy of the higher order WKB approach, it has been recently proposed to use Pad\'{e} approximation \cite{Matyjasek:2017psv, Matyjasek:2019eeu} on the usual WKB formula.These works show that Pad\'{e} approximations often works well even beyond the range of applicability of WKB approximation. In order to understand whether it is possible to construct a better approximation to achieve more accurate results, it was
	found out that by extending the order of the WKB terms i.e. increasing the order of the Taylor series approximation of the potential and constructing the well know Pad\'{e} approximants of the formal series for $\omega^2$, the Pad\'{e} transforms are always in good agreement with the exact numerically obtained QNMs.  
	
	Here in this paper we have generated the quasinomal frequencies using the 3rd order WKB and Pad\'{e} approximation and quoted both results in order to look for the improvements that the Pad\'{e} approximation induces. 
	\section{Results}
	We have numerically obtained the quasinormal frequencies for the scalar, electromagnetic and Dirac perturbations. We have exploited the 3rd order WKB approximation and Pad\'{e} approximation for calculating the frequencies of the  four dimensional Einstein-Gauss-Bonnet de Sitter black hole.
	The frequencies have been obtained for a wide range of parameter values, by individually varying $l$, $\alpha$ and $\Lambda$. The results for all three types of perturbations are presented in Table \ref{tab:qnf1} and Table \ref{tab:qnf2}. Our findings are summarised in Fig. (\ref{fig:qnf}),(\ref{fig:qnf1}).
	\subsection{Scalar perturbation}
	The inference that can be made from the above figures is that both the oscillation frequency and the damping rate decreases with increasing values of $\Lambda$. Also as $\alpha$  decreases and eventually becomes negative, the real part of the frequency decreases whereas the imaginary part becomes more negative implying that the damping increases. For positive increasing values of alpha the real part of the frequency increases, except for the $l=0$ mode where the real part was found to be decreasing with increasing $\alpha$, whereas the imaginary part increases. 
	\subsection{Electromagnetic perturbation}
	We observed similar behaviour to the case of scalar perturbation in case of the electromagnetic perturbation i.e. both the oscillation frequency and the damping rate decreases with increasing values of $\Lambda$. As $\alpha$ becomes more negative, the real part of the frequency decreases whereas the imaginary part becomes increasingly more negative implying that the decay of the modes is faster. For increasing values of $\alpha$, both the real part and the imaginary part increases, implying increasing oscillation and a decrease in the damping rate.
	\begin{figure*}[!htbp]
		\centering
		\subfloat[Scalar perturbation($\alpha$ is varied)]{{\includegraphics[width=5.43cm]{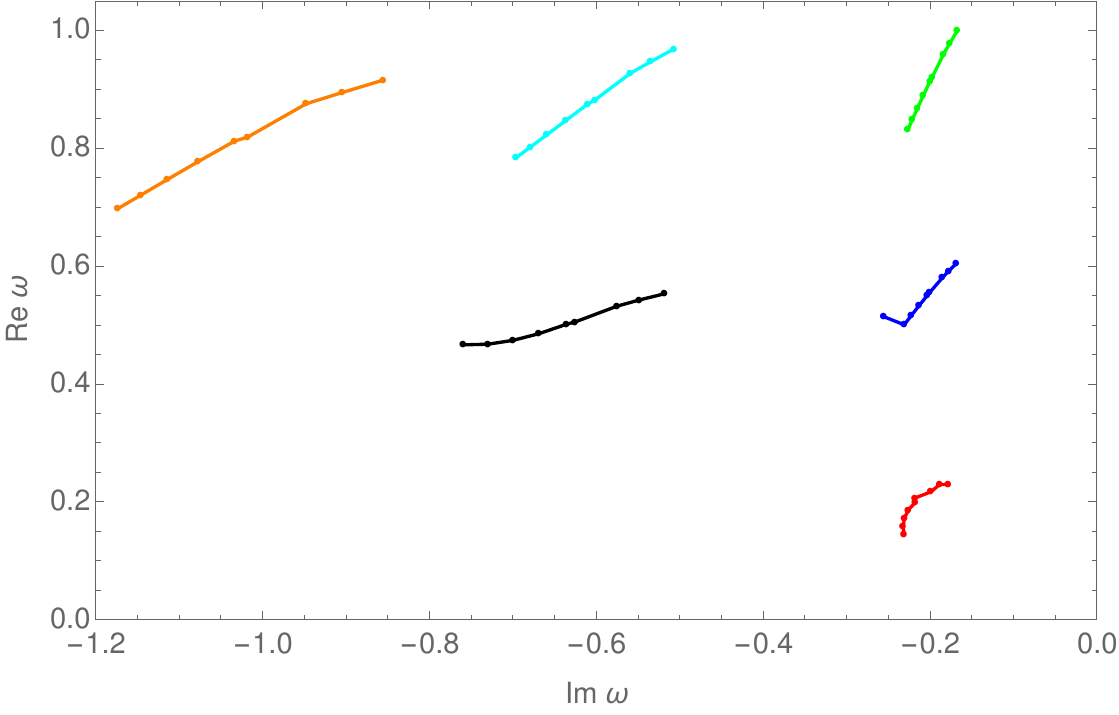} }}%
		\qquad
		\subfloat[Scalar perturbation($\Lambda$ is varied)]{{\includegraphics[width=5.43cm]{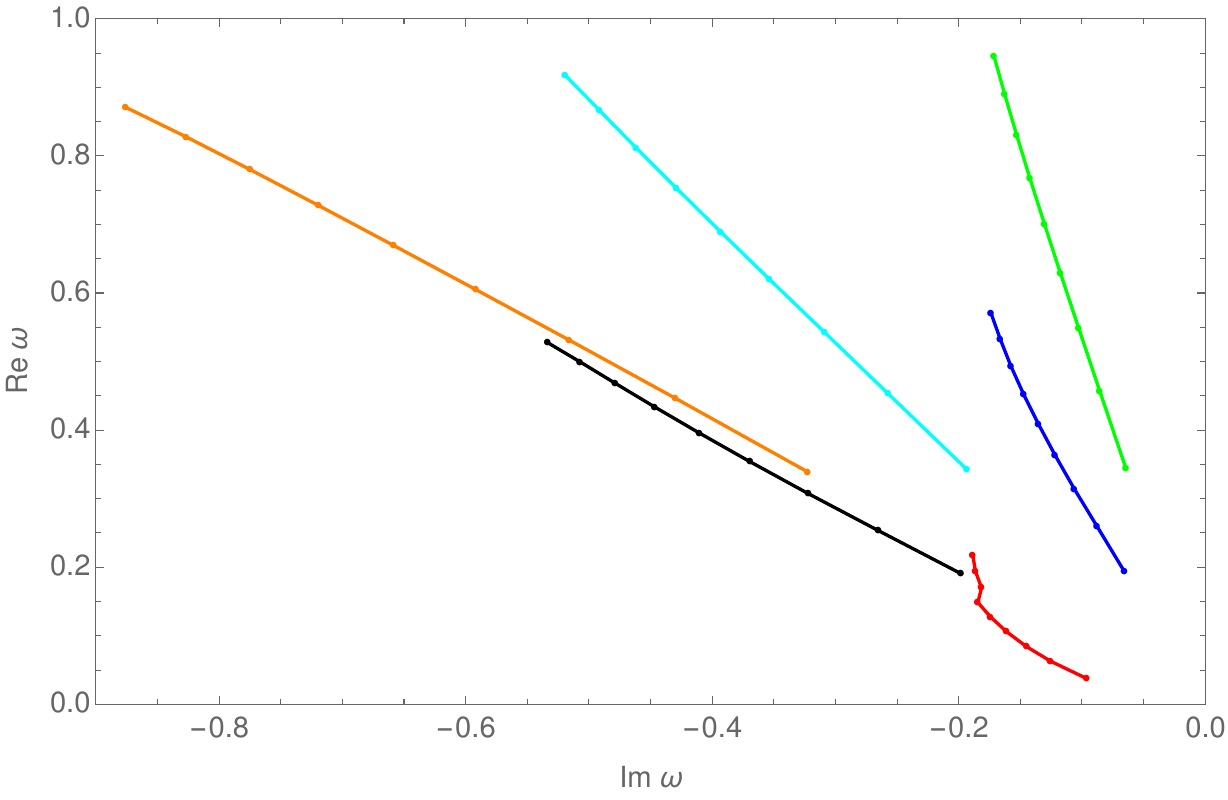} }}%
		\qquad
		\subfloat[Electromagnetic perturbation($\alpha$ is varied)]{{\includegraphics[width=5.43cm]{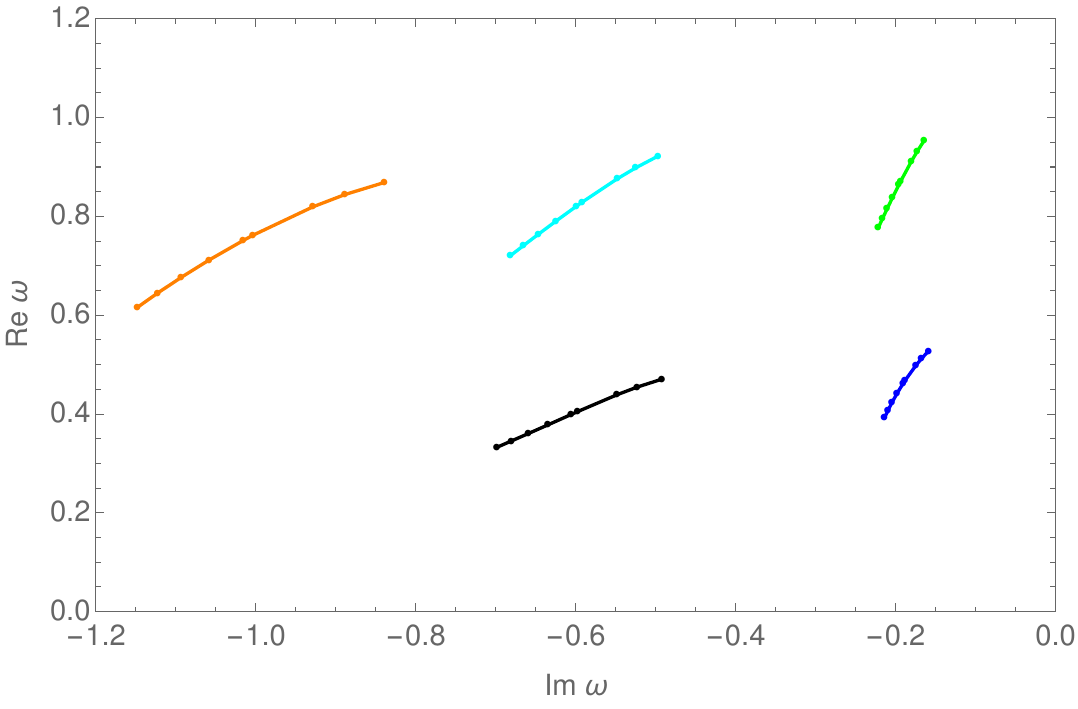} }}%
		\qquad
		\subfloat[Electromagnetic perturbation($\Lambda$ is varied)]{{\includegraphics[width=5.43cm]{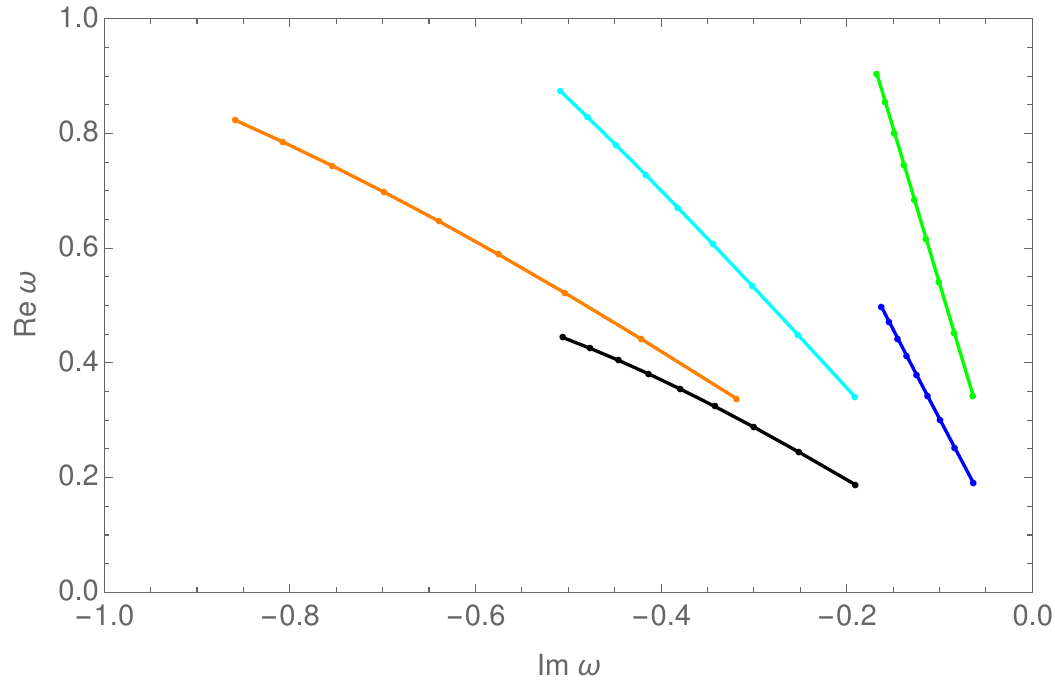} }}%
		\qquad
		\subfloat[Dirac perturbation($\alpha$ is varied)]{{\includegraphics[width=5.43cm]{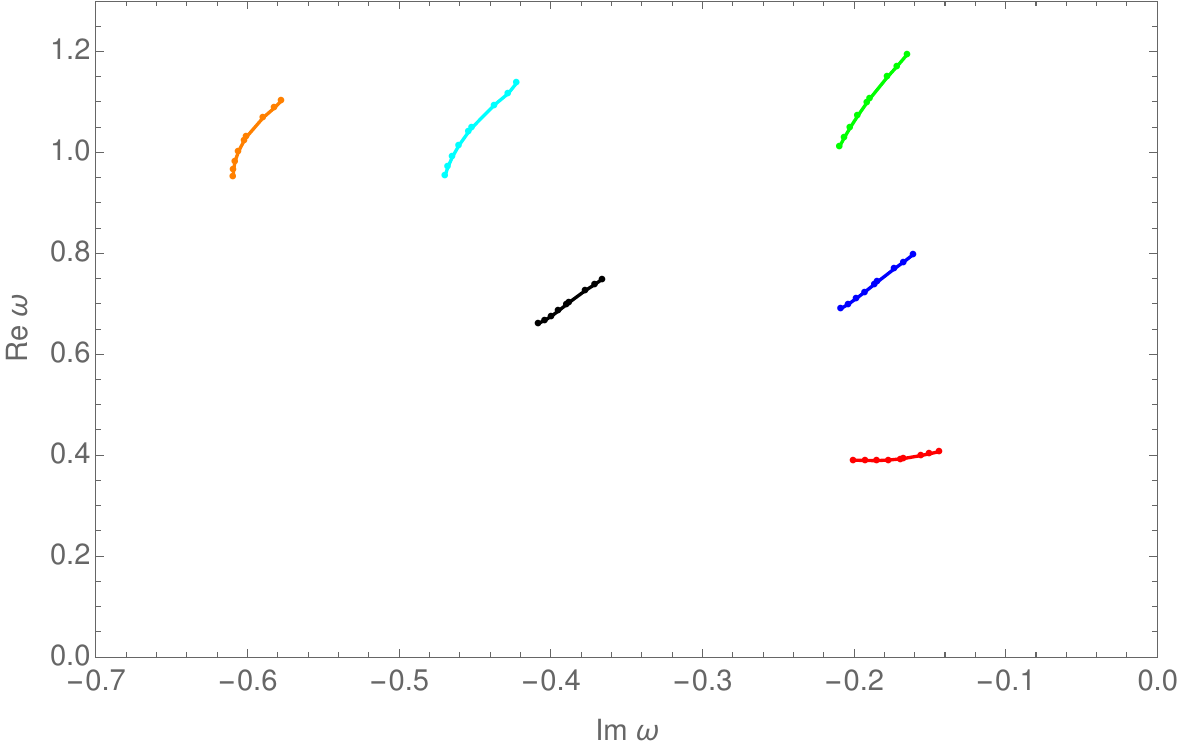} }}%
		\qquad
		\subfloat[Dirac perturbation($\Lambda$ is varied)]{{\includegraphics[width=5.43cm]{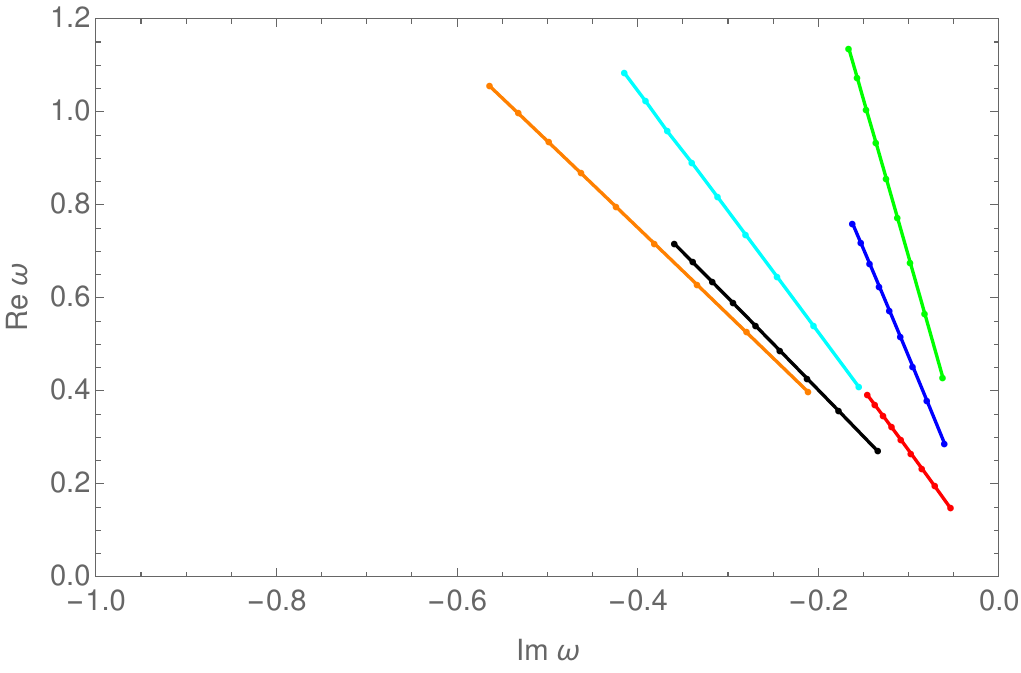} }}%
		\caption{The figure plots the real  part of the quasinormal frequencies vs the imaginary part for different parameters $(\alpha,\Lambda)$ of the four dimensional Gauss-Bonnet de Sitter spacetime, for different types of perturbations, for a fixed mass of $M=0.5$. The different colors denote the different modes for different values of $(l,n)$: red $(0,0)$; blue $(1,0)$; black $(1,1)$; green $(2,0)$; cyan $(2,1)$ and orange $(2,2)$}%
		\label{fig:qnf1}%
	\end{figure*}
	\subsection{Dirac perturbation}
	The qualitative behaviour of the Dirac case is also pretty similar to the above two cases. The oscillation frequency and the damping rate decreases with increasing values of $\Lambda$.
	With decreasing values of $\alpha$, the real part of the frequency decreases whereas the imaginary part becomes more negative implying increase in the damping rate and vice-versa. One common nature observed in all the three cases is that for a fixed value of $l$, as $n$ increases, the real part of the frequency decreases whereas the imaginary part becomes more negative implying that as the overtone increases for a fixed $l$, the damping rate increases.
	We also observe that for all the three types of perturbations, as $\alpha$ decreases,the oscillation frequency decreases with increasing damping rates whereas as $\Lambda$ decreases,the oscillation frequency increases with increasing damping rates.The results are summarised in Fig. (\ref{fig:qnf})and Fig. (\ref{fig:qnf1}), where we have plotted the real and imaginary part of the quasinormal frequency for different sets of parameter values. In particular, the plot of real vs. imaginary part of the frequencies tell us the same story as we have observed above: for all three types of perturbations studied in this work, the behaviours are quite similar for the real vs. imaginary parts of $\omega$ when we (i) varied $\alpha$, keeping $\Lambda$ fixed and (ii) varied $\Lambda$, keeping $\alpha$ fixed. In the case (i), the ${\rm{Re}}(\omega)$ grows as $-{\rm{Im}}(\omega)$ decreases while for the case (ii),  ${\rm{Re}}(\omega)$ decreases as $-{\rm{Im}}(\omega)$ decreases. 
	\begin{table*}
		\centering
		\scalebox{0.9}
		{\renewcommand{\arraystretch}{1.2}
			\begin{tabular}{|p{0.5cm}|c|c|c|c|c|c|}
				\hline
				\multirow{2}{*}{$\Lambda$} & \multicolumn{2}{c|}{\textit{Scalar frequencies}} & \multicolumn{2}{c|}{\textit{EM frequencies}} & \multicolumn{2}{c|}{\textit{Dirac frequencies}}\\
				\cline{2-7}
				& WKB & Pad\'{e} & WKB & Pad\'{e} & WKB & Pad\'{e}\\
				\hline
				\multicolumn{7}{|c|}{$l=0$, $n=0$ }\\
				\hline
				0.05 & 0.191851 - 0.203018 i &0.217115 - 0.188491 i & - & - &  0.35688 - 0.173238 i &0.39021 - 0.14534 i\\
				\hline
				0.1 &0.176822 - 0.200981  i&0.194184-0.186394 i & -&- & 0.338482 - 0.162822 i&0.368328-0.13681 i\\
				\hline
				0.15 &  0.160047 - 0.196268i &0.169978-0.181467 i & - & - &  0.318778 - 0.151993 i&0.34516-0.127841 i\\
				\hline
				0.2 &  0.141785 - 0.18859 i &0.148664-0.184423 i & - & - & 0.297486 - 0.140628 i&0.320425-0.118332 i\\
				\hline
				0.25 & 0.122442 - 0.17803 i &0.127506-0.174134 i & - & - & 0.274207 - 0.12855 i&0.293732-0.108145 i\\
				\hline
				\multicolumn{7}{|c|}{ $l=1$, $n=0$}\\
				\hline
				0.05 & 0.565756 - 0.17396 i & 0.569998 - 0.173652 i & 0.491597 - 0.162554 i & 0.497129 - 0.162263 i & 0.745408 - 0.169465 i & 0.758533 - 0.16181 i\\
				\hline
				0.1 & 0.528726 - 0.166532 i & 0.532142-0.166151 i & 0.46572 - 0.154036 i & 0.470431-0.153756 i & 0.704743 - 0.159868 i & 0.71612-0.152461 i\\
				\hline
				0.15 & 0.490219 - 0.157751 i & 0.492718-0.157472 i & 0.438005 - 0.144944 i & 0.441911-0.14467 i & 0.661493 - 0.149753 i & 0.671201-0.142623 i\\
				\hline
				0.2 & 0.450008 - 0.147533 i & 0.451764-0.147167 i & 0.408084 - 0.135147 i & 0.411209-0.134873 i & 0.615113 - 0.138992 i & 0.623236-0.132176 i\\
				\hline
				0.25 & 0.407704 - 0.135771 i & 0.408856-0.135299 i & 0.375435 - 0.124459 i & 0.377818-0.124186 i & 0.564831 - 0.127412 i & 0.571455-0.120958 i\\
				\hline
				
				\multicolumn{7}{|c|}{$l=1$, $n=1$}\\
				\hline
				0.05 & 0.52127 - 0.53719 i & 0.527453 - 0.533341 i & 0.435811 - 0.510813 i & 0.444061 - 0.505422 i & 0.708779 - 0.51868 i & 0.715535 - 0.359014 i\\
				\hline
				0.1 & 0.494656 - 0.510235 i & 0.49951-0.50712 i & 0.418053 - 0.480494 i & 0.425114-0.47587 i & 0.673411 - 0.487725 i & 0.675567-0.338491 i\\
				\hline
				0.15 & 0.464595 - 0.480984 i & 0.46831-0.478612 i & 0.398227 - 0.449089 i & 0.404096-0.445216 i & 0.635276 - 0.455507 i & 0.633195-0.316863 i\\
				\hline
				0.2 & 0.431015 - 0.448461 i & 0.433718-0.446559 i & 0.375867 - 0.416157 i & 0.380581-0.413006 i & 0.593788 - 0.421635 i & 0.587946-0.293851 i\\
				\hline
				0.25 & 0.393812 - 0.411944 i & 0.395598-0.41023 i & 0.350355 - 0.381107 i & 0.353984-0.378636 i & 0.54813 - 0.385565 i & 0.539103-0.269094 i\\
				\hline
				\multicolumn{7}{|c|}{$l=2$, $n=0$}\\
				\hline
				0.05 & 0.944699 - 0.1714 i & 0.945696 - 0.171379 i & 0.902054 - 0.167263 i & 0.903127 - 0.167234 i & 1.12689 - 0.169579 i & 1.13428 - 0.166011 i\\
				\hline
				0.1 & 0.888283 - 0.16259 i & 0.889111-0.162562 i & 0.852451 - 0.158068 i & 0.853368-0.15804 i & 1.06456 - 0.160023 i & 1.07089-0.15655 i\\
				\hline
				0.15 & 0.829162 - 0.152897 i & 0.829823-0.152864 i & 0.799758 - 0.148314 i & 0.80052-0.148286 i & 0.998434 - 0.149928 i & 1.00374-0.146701 i\\
				\hline
				0.2 & 0.766736 - 0.142233 i & 0.767245-0.142194 i & 0.743323 - 0.137876 i & 0.743934-0.137848 i & 0.927684 - 0.13917 i & 0.932051-0.135917 i\\
				\hline
				0.25 & 0.700131 - 0.130461 i & 0.700513-0.130413 i & 0.68222 - 0.126579 i & 0.682688-0.126551 i & 0.851162 - 0.127577 i& 0.854649-0.124474 i\\
				\hline
				\multicolumn{7}{|c|}{$l=2$, $n=1$}\\
				\hline
				0.05 & 0.915765 - 0.519938 i & 0.917682 - 0.519229 i & 0.870677 - 0.508523 i & 0.87277 - 0.507717 i & 1.10179 - 0.513102 i & 1.08211 - 0.414238 i\\
				\hline
				0.1 & 0.864759 - 0.492051 i & 0.866335-0.491433 i & 0.825968 - 0.479385 i & 0.827758-0.478689 i & 1.04337 - 0.483418 i & 1.02162-0.390803 i\\
				\hline
				0.15 & 0.810215 - 0.461984 i & 0.811479-0.461475 i & 0.777896 - 0.448823 i & 0.779385-0.448236 i & 0.980924 - 0.452281 i & 0.95681-0.366767 i\\
				\hline
				0.2 & 0.751675 - 0.429274 i & 0.752663-0.428857 i & 0.779385-0.448236 i & 0.726982-0.415958 i & 0.913617 - 0.419307 i & 0.88925-0.339532 i\\
				\hline
				0.25 & 0.688381 - 0.393375 i & 0.689123-0.392995 i & 0.668676 - 0.381683 i & 0.669603-0.381303 i & 0.84028 - 0.383964 i & 0.815449-0.311051 i\\
				\hline
				\multicolumn{7}{|c|}{$l=2$, $n=2$}\\
				\hline
				0.05 & 0.866557 - 0.878753 i & 0.870839 - 0.875728 i & 0.818191 - 0.861649 i & 0.822857 - 0.858214 i & 1.05787 - 0.865609 i & 1.05469 - 0.563642 i\\
				\hline
				0.1 & 0.824142 - 0.829065 i & 0.827575-0.826545 i & 0.780821 - 0.809826 i & 0.784814-0.806893 i & 1.00575 - 0.813699 i & 0.995787-0.531736 i\\
				\hline
				0.15 & 0.777282 - 0.77677 i & 0.779939-0.774784 i & 0.739979 - 0.756148 i & 0.743273-0.753722 i & 0.949425 - 0.759757 i & 0.933381-0.49803 i\\
				\hline
				0.2 & 0.725457 - 0.720678 i & 0.727435-0.719142 i & 0.69488 - 0.699911 i & 0.697483-0.697981 i & 0.888002 - 0.703117 i & 0.866721-0.462122 i\\
				\hline
				0.25 & 0.667922 - 0.659599 i & 0.669273-0.658321 i & 0.644467 - 0.640159 i & 0.646417-0.638696 i & 0.82024 - 0.642859 i & 0.794751-0.423441 i\\
				\hline
		\end{tabular}}
		\caption{The table shows the quasinormal frequencies for scalar, electromagnetic and Dirac perturbation calculated using 3rd order WKB and 3rd order Pad\'{e} approximation for different modes and for different values of the cosmological constant $\Lambda$ with a fixed value of $\alpha=0.2$} 
		\label{tab:qnf1}
	\end{table*}
	\begin{table*}
		\scalebox{0.9}
		{\renewcommand{\arraystretch}{1.2}
			\begin{tabular}{|p{0.8cm}|c|c|c|c|c|c|}
				\hline
				\multirow{2}{*}{$\alpha$} & \multicolumn{2}{c|}{\textit{Scalar frequencies}} & \multicolumn{2}{c|}{\textit{EM frequencies}} & \multicolumn{2}{c|}{\textit{Dirac frequencies}}\\
				\cline{2-7}
				& WKB & Pad\'{e} & WKB & Pad\'{e} & WKB & Pad\'{e}\\
				\hline
				\multicolumn{7}{|c|}{ $l=0$, $n=0$}\\
				\hline
				-0.4& 0.180832 - 0.27735 i & 0.186337-0.225981 i & - & - & 0.296146 - 0.219785 i & 0.390391-0.177111 i\\
				\hline
				-0.2 & 0.195772 - 0.253448 i & 0.198904-0.217768 i & - & - & 0.322481 - 0.208132 i & 0.392702-0.169171 i\\
				\hline
				0.1 & 0.201651 - 0.214774 i & 0.217165-0.198971 i & - & - & 0.35669 - 0.187761 i & 0.399369-0.155677 i\\
				\hline
				0.2 & 0.199762 - 0.203244 i & 0.229576-0.188228 i & - & - & 0.367382 - 0.179329 i & 0.402809-0.150239 i\\
				\hline
				0.3 & 0.193427 - 0.190947 i & 0.229452-0.178089 i & - & - & 0.377815 - 0.169283 i & 0.40716-0.143584 i\\
				\hline
				\multicolumn{7}{|c|}{ $l=1$, $n=0$}\\
				\hline
				-0.4& 0.536686 - 0.218129 i & 0.532734-0.213291 i & 0.446086 - 0.205134 i & 0.442128-0.197707 i & 0.699842 - 0.209979 i & 0.72356-0.192807 i\\
				\hline
				-0.2 & 0.549937 - 0.206004 i & 0.550216-0.203365 i & 0.461665 - 0.194293 i & 0.462492-0.189986 i & 0.717943 - 0.1996 i & 0.739206-0.186294 i\\
				\hline
				0.1 & 0.57638 - 0.185835 i& 0.580238-0.185338 i & 0.493014 - 0.175219 i & 0.498365-0.174221 i & 0.753786 - 0.182295 i & 0.769883-0.173194 i\\
				\hline
				0.2 & 0.587311 - 0.177812 i & 0.591926-0.177599 i & 0.506358 - 0.167434 i & 0.512385-0.167129 i & 0.768739 - 0.175014 i & 0.782949-0.167197 i\\
				\hline
				0.3 & 0.599501 - 0.168454 i & 0.604401-0.16845 i & 0.521617 - 0.1582 i & 0.527825-0.158251 i & 0.78572 - 0.166146 i & 0.798141-0.160778 i\\
				\hline
				\multicolumn{7}{|c|}{$l=1$, $n=1$}\\
				\hline
				-0.4& 0.47996 - 0.673781 i & 0.485412-0.668869 i & 0.36924 - 0.643044 i & 0.378135-0.634388 i & 0.63944 - 0.652996 i & 0.686091-0.394849 i\\
				\hline
				-0.2 & 0.495205 - 0.640069 i & 0.501489-0.635617 i & 0.39027 - 0.613196 i & 0.399185-0.605314 i & 0.662923 - 0.619354 i & 0.699135-0.389366 i\\
				\hline
				0.1 & 0.524966 - 0.578625 i & 0.532149-0.575213 i & 0.430211 - 0.554447 i & 0.439029-0.548075 i & 0.710442 - 0.561208 i & 0.726343-0.377132 i\\
				\hline
				0.2 & 0.535675 - 0.552912 i & 0.542598-0.548649 i & 0.44563 - 0.528629 i & 0.454575-0.522776 i & 0.72886 - 0.536773 i & 0.738618-0.370809 i\\
				\hline
				0.3 & 0.545344 - 0.522544 i & 0.553166-0.518376 i & 0.461115 - 0.49729 i & 0.470598-0.491876 i & 0.747842 - 0.507336 i & 0.749214-0.365865 i\\
				\hline
				\multicolumn{7}{|c|}{$l=2$, $n=0$}\\
				\hline
				-0.4& 0.890563 - 0.20879 i & 0.889549-0.208055 i & 0.839297 - 0.20436 i & 0.838182-0.203475 i & 1.06099 - 0.206307 i & 1.07246-0.19753 i\\
				\hline
				-0.2 & 0.914376 - 0.200167 i & 0.914363-0.19977 i & 0.864146 - 0.196003 i & 0.864129-0.195497 i & 1.08844 - 0.198134 i & 1.09919-0.191307 i\\
				\hline
				0.1 & 0.959097 - 0.183588 i & 0.959925-0.183664 i & 0.911108 - 0.179739 i & 0.912119-0.179628 i & 1.14091 - 0.182208 i & 1.14975-0.177852 i\\
				\hline
				0.2 & 0.977433 - 0.17629 i & 0.978527-0.176265 i & 0.930539 - 0.172552 i & 0.931706-0.172521 i & 1.16271 - 0.175093 i & 1.17078-0.171533 i\\
				\hline
				0.3 & 0.99823 - 0.167449 i & 0.999374-0.167457 i & 0.952729 - 0.163741 i & 0.953937-0.163749 i & 1.18763 - 0.166321 i & 1.19494-0.164708 i\\
				\hline
				\multicolumn{7}{|c|}{$l=2$, $n=1$}\\
				\hline
				-0.4& 0.844874 - 0.637983 i & 0.847039-0.636781 i & 0.787721 - 0.626041 i & 0.790421-0.624512 i & 1.01882 - 0.629338 i & 1.01416-0.460468 i\\
				\hline
				-0.2& 0.872879 - 0.611238 i & 0.875131-0.610292 i & 0.818125 - 0.600068 i & 0.820701-0.598741 i & 1.05027 - 0.603375 i & 1.04124-0.454037 i\\
				\hline
				0.1 & 0.924658 - 0.559073 i & 0.927215-0.559376 i & 0.874034 - 0.548465 i & 0.876359-0.547481 i & 1.11083 - 0.552839 i & 1.0941-0.437141 i\\
				\hline
				0.2 & 0.944793 - 0.535835 i & 0.946892-0.535036 i & 0.896104 - 0.525452 i & 0.898376-0.52458 i & 1.13518 - 0.530337 i & 1.11664-0.428069 i\\
				\hline
				0.3 & 0.966438 - 0.507735 i & 0.968645-0.507014 i & 0.920002 - 0.497242 i & 0.922308-0.496468 i & 1.16186 - 0.502703 i & 1.13789-0.422383 i\\
				\hline
				\multicolumn{7}{|c|}{$l=2$, $n=2$}\\
				\hline
				-0.4& 0.771374 - 1.08336 i & 0.777263-1.07762 i & 0.704299 - 1.06566 i & 0.711909-1.05811 i & 0.950708 - 1.06974 i & 1.00201-0.605908 i\\
				\hline
				-0.2 & 0.806628 - 1.03816 i & 0.812274-1.03368 i & 0.744812 - 1.0216 i & 0.751293-1.01558 i & 0.987769 - 1.02449 i & 1.0245-0.601971 i\\
				\hline
				0.1 & 0.868698 - 0.94851 i & 0.875572-0.947909 i & 0.814313 - 0.932399 i & 0.819472-0.928275 i & 1.05991 - 0.935659 i & 1.06913-0.58956 i\\
				\hline
				0.2 & 0.890101 - 0.907925 i & 0.894772-0.904524 i & 0.839203 - 0.892084 i & 0.844248-0.888357 i & 1.08741 - 0.895997 i & 1.08845-0.582131 i\\
				\hline
				0.3 & 0.910593 - 0.858723 i & 0.915856-0.85535 i & 0.863356 - 0.842427 i & 0.868766-0.838855 i & 1.11535 - 0.847334 i & 1.10249-0.577596 i\\
				\hline
		\end{tabular}}
		\caption{The table shows the quasinormal frequencies for scalar, electromagnetic and Dirac perturbation calculated using 3rd order WKB and 3rd order Pad\'{e} approximation for different modes and for different values of the coupling constant $\alpha$ with a fixed value of $\Lambda=0.02$} 
		\label{tab:qnf2}
	\end{table*}
	\section{Eikonal QNMs, Lyapunov exponents and null geodesics}
	In the previous section we studied the quasi-normal frequencies for the four dimensional Gauss-Bonnet deSitter black hole solution employing third order WKB and Pad\'{e} approximants. In this section, we would be interested in looking at the QNFs in the eikonal limit i.e. for very very large $l$ values.
	
	It has been well known that for static spacetimes, the scalar, electromagnetic and Dirac perturbations have similar behaviour in the eikonal limit \cite{Kodama:2003kk} and their effective potential in this limit could be simultaneously given by 
	\begin{equation}
	V_{eikonal}=l(l+1)\frac{f(r)}{r^2}
	\end{equation}
	Exploiting this simple observation and the fact that the peak of the effective potentials in the eikonal limit, $r_0$, coincides with that of the unstable null geodesics $r_p$, Cardoso \textit{et.al.} \cite{Cardoso:2008bp} showed that the QNFs of a spherically symmetric, asymptotically non-AdS black hole, in the eikonal limit could be expressed by a very simple formula, which only depends on the metric function $f(r)$ and the position of the unstable null geodesic $r_p$, given by 
	\begin{equation}\label{eq:eikonal}
	\omega_{QNM}=\Omega_p l - i(n+1/2)|\lambda|
	\end{equation}
	where, $\Omega_p=\sqrt{\frac{f_p}{r_p^2}}$ and 
	\begin{equation}
	\lambda=\frac{1}{\sqrt{2}}\sqrt{\frac{-r_p^2}{f_p}\Bigg(\frac{d^2}{dr_*^2}\frac{f}{r^2}\Bigg)_{r=r_p}}
	\end{equation}
	where the subscript $p$ denotes that the corresponding quantity has been calculated at the unstable null radius $r_p$ and $r_*$ is the tortoise coordinate. Physically, $\Omega_p$ denotes the angular frequency of the unstable orbiting photons and $\lambda$ denotes the principle Lyapunov exponent at the unstable null geodesics. We find the eikonal frequency using the third order WKB approximation for all three type of perturbation and from the approximate formula Eq. (\ref{eq:eikonal}) and report the numbers in Table \ref{tab:eikonal}. The convergence of the frequencies with each other and with the approximate formula with the increasing value of $l$ is evident from the table.
	\begin{table*}
		{
			\begin{tabular}{|c|c|c|c|c|}
				\hline
				$l$& $Scalar$ &$EM$&$Dirac$&$Approximate$ $frequency$ $from$ $Eq.$ (\ref{eq:eikonal})\\
				\hline
				3500 & 1321.97 - 0.169845 \textbf{\textit{i}}&1321.97 - 0.169845 \textbf{\textit{i}}&1322.16 - 0.169845 \textbf{\textit{i}} & 1321.778513516667734 - 0.169844849949930035 \textbf{\textit{i}}\\
				\hline
				4000 & 1510.79 - 0.169845 \textbf{\textit{i}}&1510.79 - 0.169845 \textbf{\textit{i}}&1510.98 - 0.169845 \textbf{\textit{i}}& 1510.604015447620313 - 0.169844849781841718 \textbf{\textit{i}}\\
				\hline
				4500 & 1699.62 - 0.169845 \textbf{\textit{i}}&1699.62 - 0.169845  \textbf{\textit{i}}&1699.81 - 0.169845 \textbf{\textit{i}}&1699.429517378572879 - 0.169844849666595429 \textbf{\textit{i}}\\
				\hline
				5000& 1888.44 - 0.169845 \textbf{\textit{i}}&1888.44 - 0.169845 \textbf{\textit{i}}&1888.63 - 0.169845 \textbf{\textit{i}} & 1888.255019309525439 - 0.169844849584157273 \textbf{\textit{i}}\\
				\hline
				5500& 2077.27 - 0.169845 \textbf{\textit{i}}&2077.27 - 0.169845  \textbf{\textit{i}}&2077.46 - 0.169845 \textbf{\textit{i}}& 2077.080521240477990 - 0.169844849523160555 \textbf{\textit{i}}\\
				\hline
			\end{tabular}
		}
		\caption{The table shows the quasinormal frequencies
			for the $n=0$ mode and for very large multipole numbers $l$ for a four dimensional Gauss-Bonnet de Sitter black hole 
			with $M=0.5$, $\alpha=0.2$ and $\Lambda=0.05$}
		\label{tab:eikonal}
	\end{table*}
	\begin{figure*}%
		\centering
		\subfloat[$\alpha=0.4$ (red) and $\alpha=0.1$ (blue) for $\Lambda=0.02$ and $l=1$ ]{{\includegraphics[width=8cm]{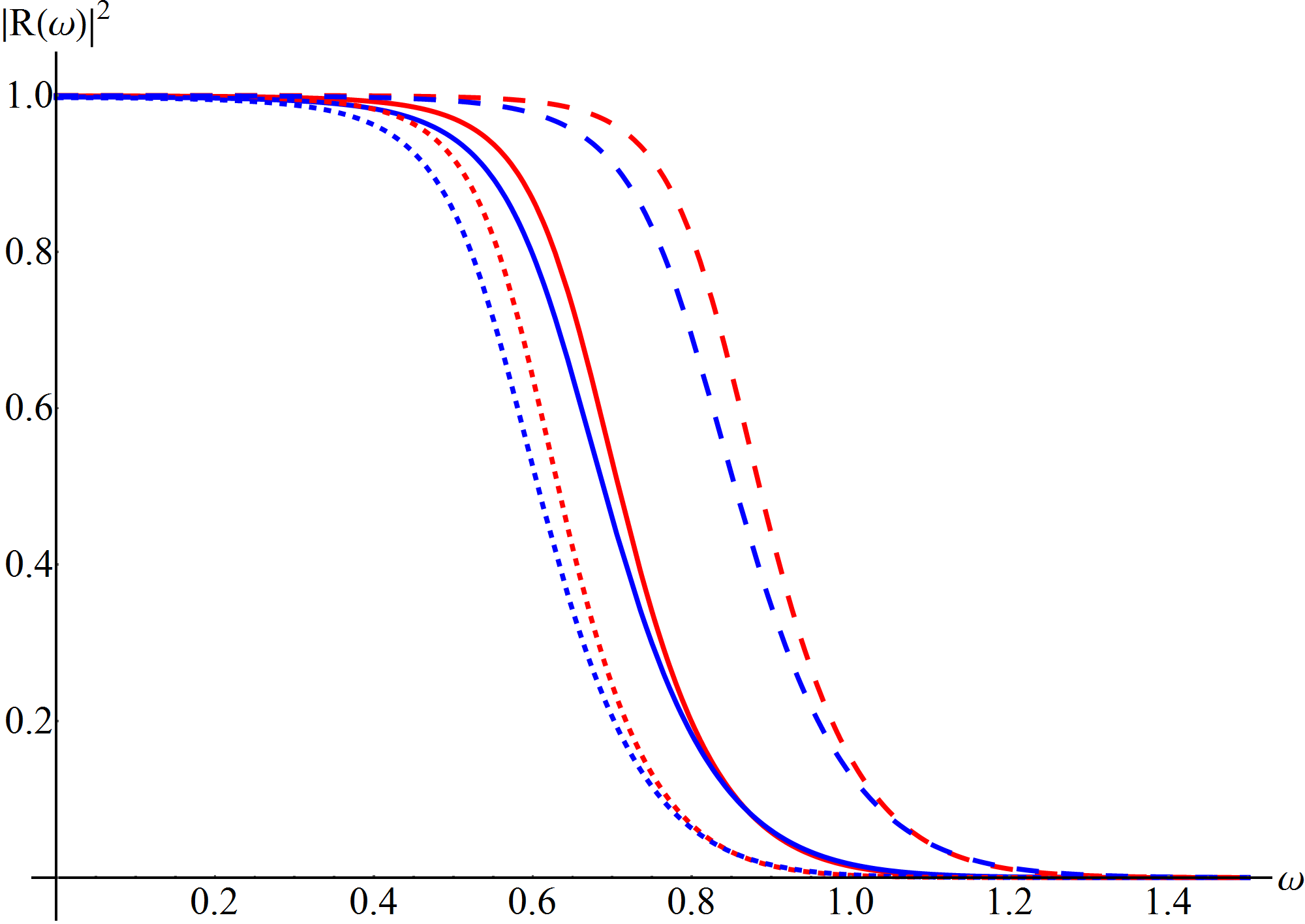} }}%
		\qquad
		\subfloat[$\alpha=0.4$ (red) and $\alpha=0.1$ (blue) for $\Lambda=0.02$ and $l=1$ ]{{\includegraphics[width=8cm]{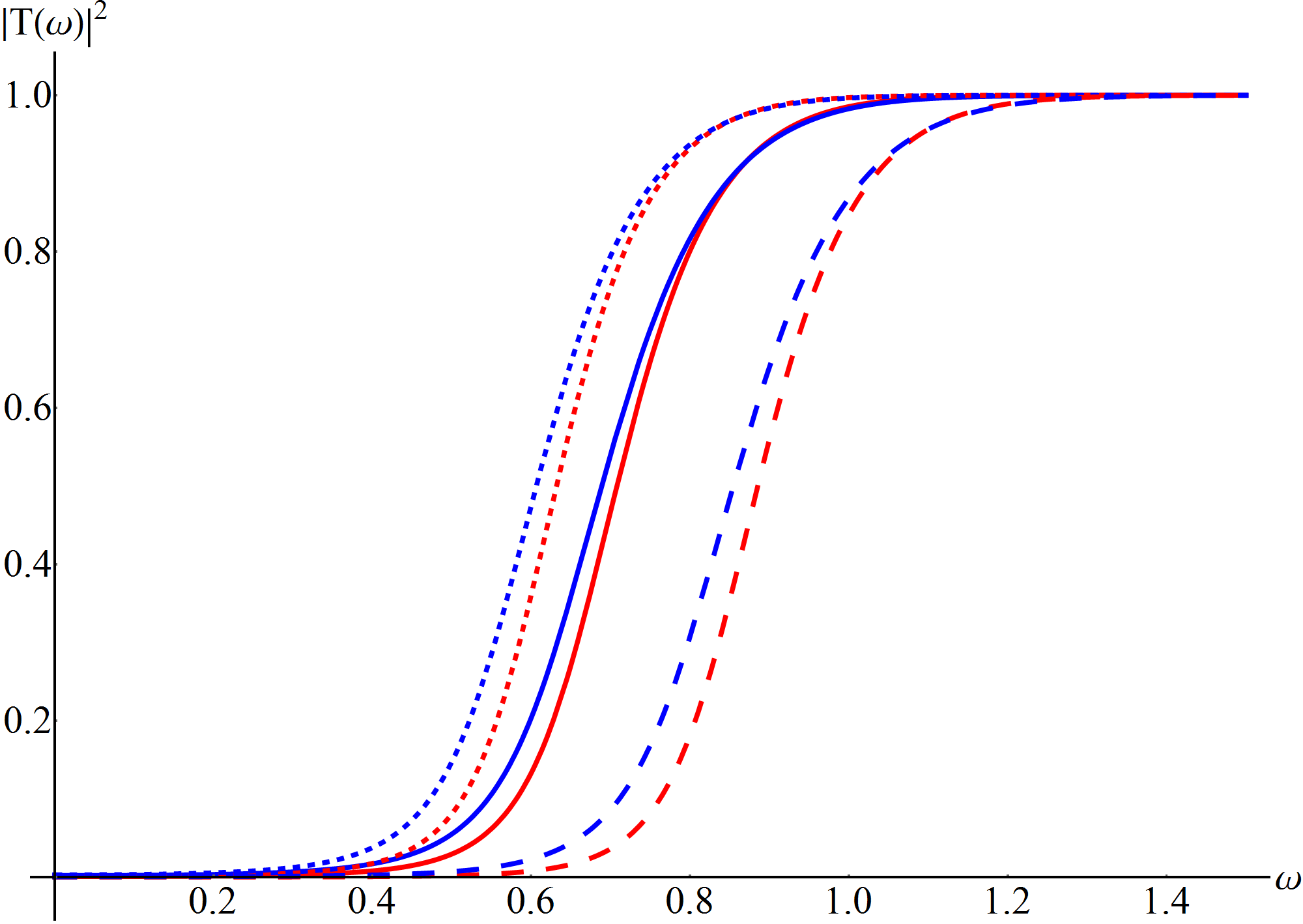} }}%
		\qquad
		\subfloat[$\Lambda=0.02$ (red) and $\Lambda=0.08$ (blue) for $\alpha=0.15$ and $l=1$]{{\includegraphics[width=8cm]{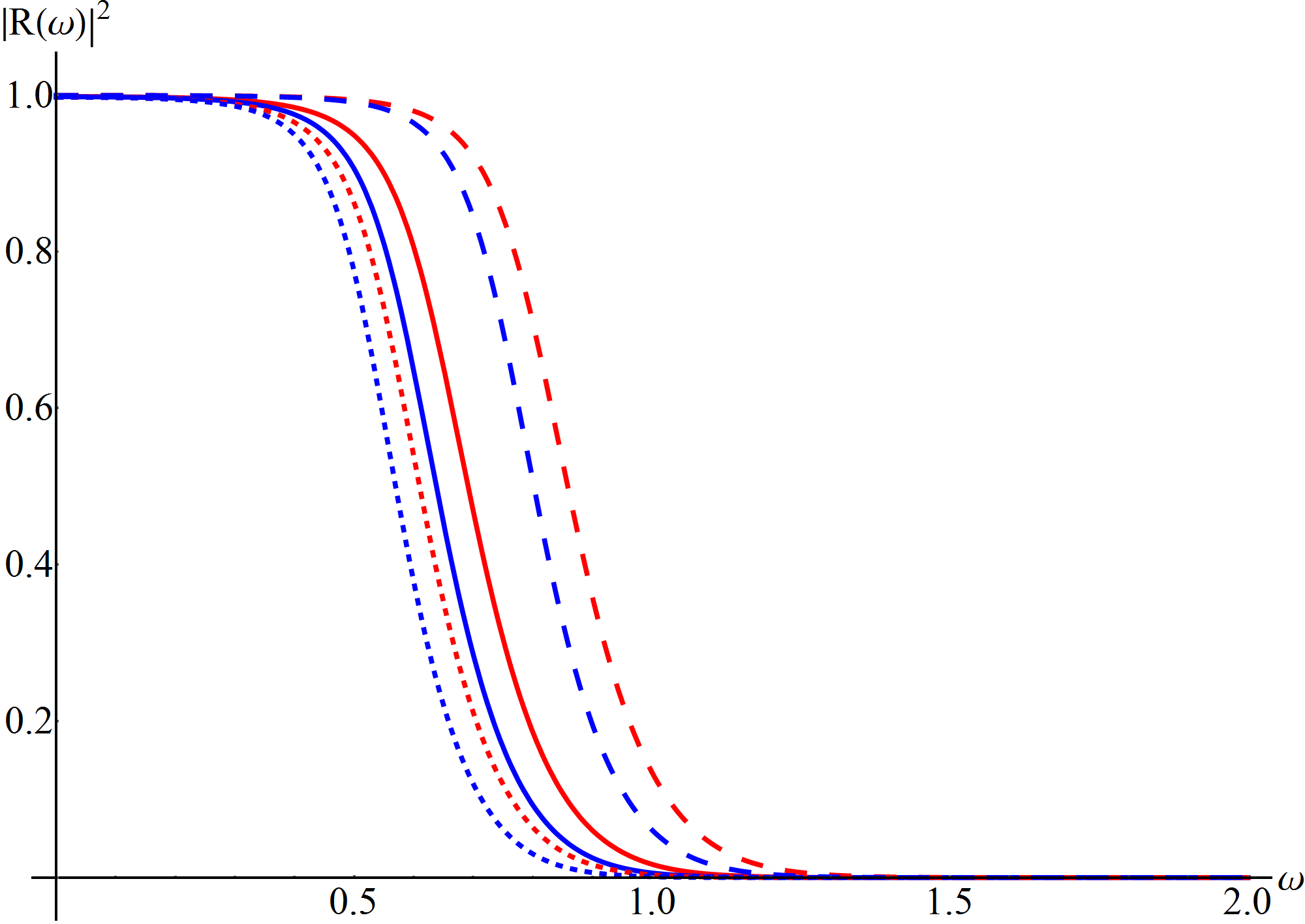} }}%
		\qquad
		\subfloat[$\Lambda=0.02$ (red) and $\Lambda=0.08$ (blue) for $\alpha=0.15$ and $l=1$]{{\includegraphics[width=8cm]{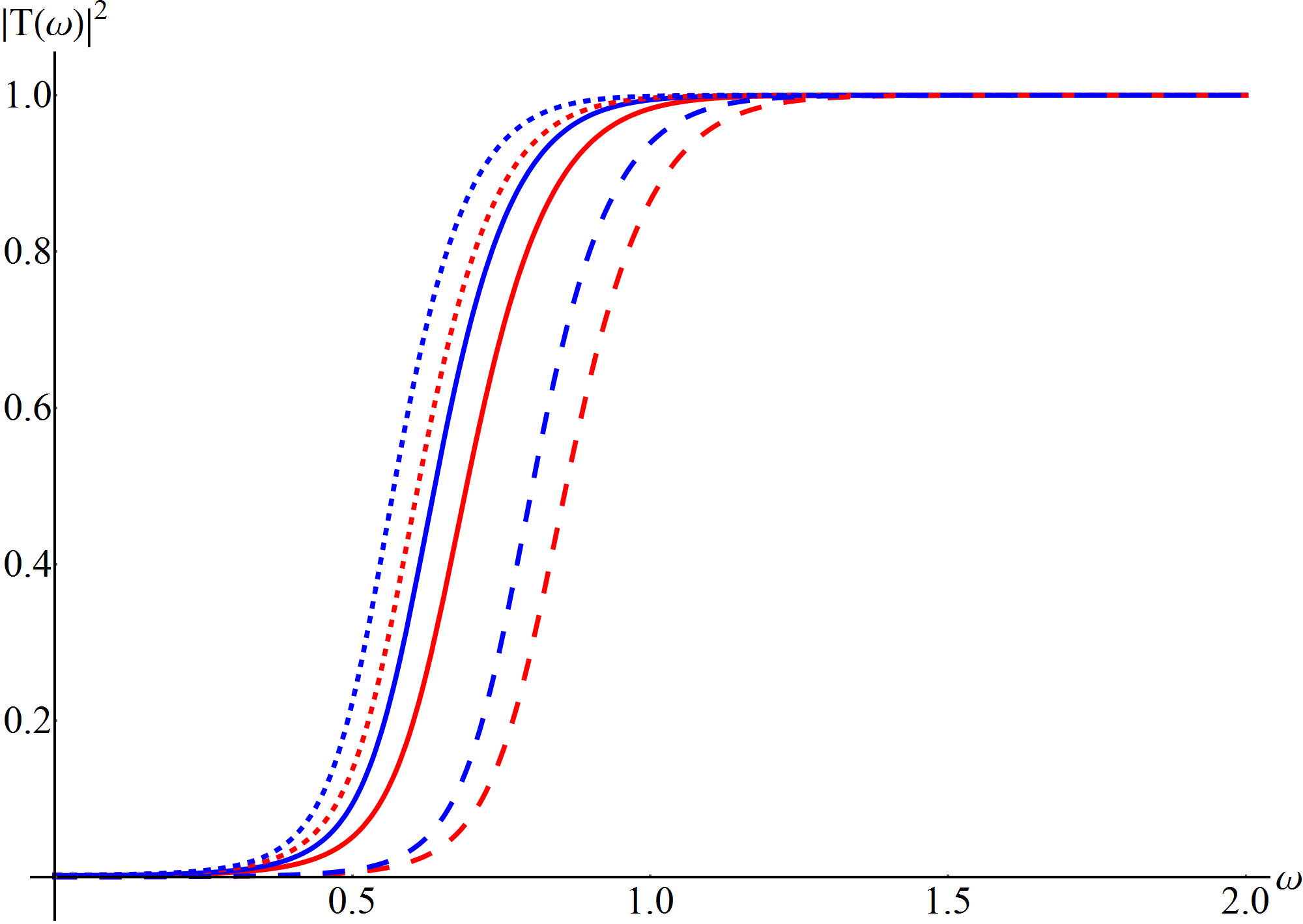} }}%
		\qquad
		\subfloat[$l=1$ (red) and $l=2$ (blue) for $\Lambda=0.02$ and $\alpha=0.15$]{{\includegraphics[width=8cm]{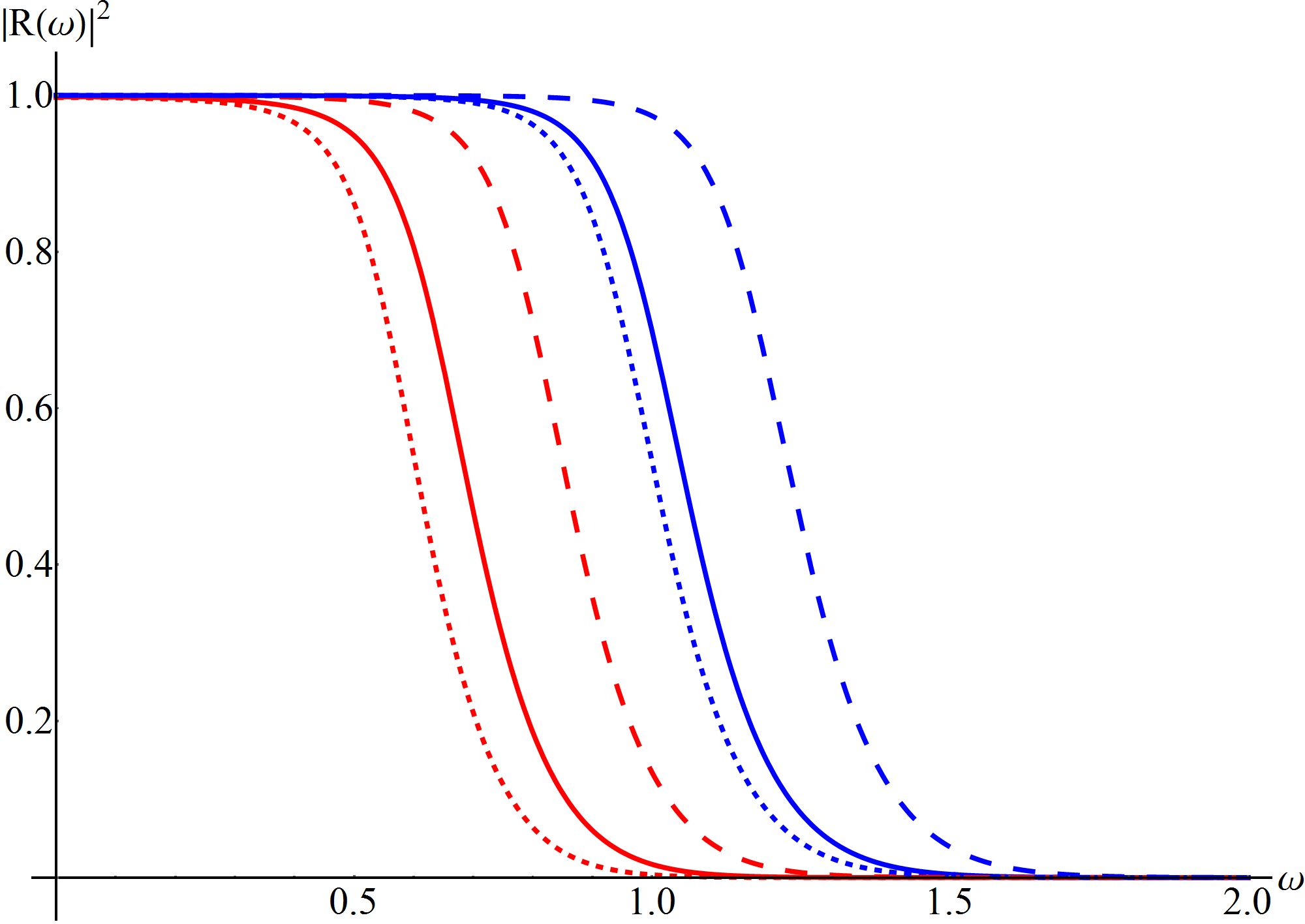} }}%
		\qquad
		\subfloat[$l=1$ (red) and $l=2$ (blue) for $\Lambda=0.02$ and $\alpha=0.15$]{{\includegraphics[width=8cm]{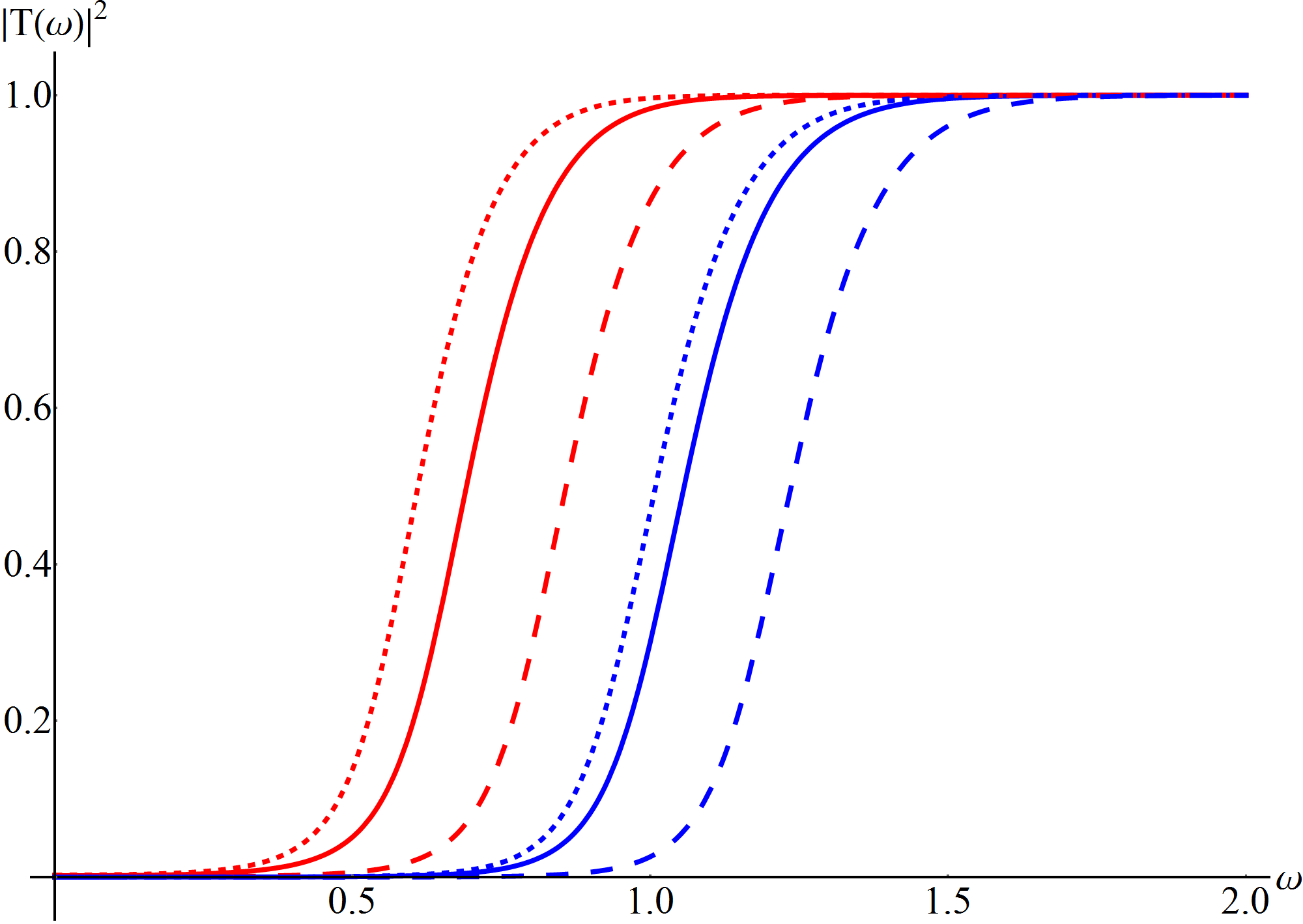} }}%
		\caption{The figure plots the reflection and transmission coefficient of the scattered scalar (solid), electromagnetic (dotted) and Dirac (dashed) wave for $M=0.5$ and different parameter values}%
		\label{fig:greybody}%
	\end{figure*}
	\section{Greybody Factor}
	In this section we discuss the frequency dependent reflection and transmission coefficient, $R(\omega)$ and $T(\omega)$ respectively, for a scattering process of the scalar, EM and Dirac wave from the black hole. This case is different from the quasi normal frequency calculation since we relax the
	boundary condition of no-incoming wave from infinity.
	After scattering off of the effective potential, the asymptotic behaviour of the wave could be written in tortoise
	coordinate as
	\begin{equation}
	\psi(r_*)=T(\omega)e^{-i\omega r_*}; r_*\rightarrow-\infty
	\end{equation}
	\begin{equation}
	\psi(r_*)=e^{-i\omega r_*}+R(\omega)e^{i\omega r_*}; r_*\rightarrow\infty
	\end{equation}
	In the WKB approximation, the reflection coefficient is
	given by
	\begin{equation}
	R(\omega)=(1+e^{-2\pi i\beta})^{\frac{1}{2}}
	\end{equation}
	where $\beta$, under the third order WKB approximation, is
	given by
	\begin{equation}
	\frac{i(\omega^2-V(r_0))}{\sqrt{-2V''(r_0)}}-V_2-V_3
	\end{equation}
	where $V_2$ and $V_3$ is given by
	\bea
	V_2&=&-i \Lambda(n),\\
	V_3&=&\frac{-1}{\sqrt{-2V_0''}}\Omega(n)
	\eea
	Conserving probability we get
	\begin{equation}
	\gamma_l=|T(\omega)|^2=1-|R(\omega)|^2
	\end{equation}
	where $\gamma_l$ is the greybody factor. This method of finding the reflection coefficient has been extensively employed in past literature. Below we plot the behaviour of the reflection and transmission coefficient, with the frequency of the wave, for a wide range of parameter values in Fig. (\ref{fig:greybody}). The general nature of the greybody factors for different types of waves is essentially similar. The greybody factors decrease with an increasing $l$ and increasing coupling constant $\alpha$. This also implies, that the greybody factors for negative $\alpha$ would be greater than positive ones. The greybody factors tend to increase with an increase in the cosmological constant. Table \ref{tab:gf} shows the behaviour of the greybody factor for all three cases for different sets of parameters $\alpha$, $\Lambda$ and $l$.
	\begin{table}
		\centering
		\scalebox{1.0}
		{\renewcommand{\arraystretch}{1.5}
			\begin{tabular}{|p{0.8cm}|c|c|c|}
				\hline
				$\omega$ &  \textit{Scalar case} &  \textit{EM case} &  \textit{Dirac case}\\
				& $|T(\omega)|^{2}$ &   $|T(\omega)|^{2}$   & $|T(\omega)|^{2}$ \\
				\hline
				\multicolumn{4}{|c|}{ $l=1$, \,\,\,\,$n=0$,\,\,\,\, $\alpha=0.4$,\,\,\,\, $\Lambda=0.02$}\\
				\hline
				0& 0.000773933&  0.00108008&   0.000104834\\
				\hline
				0.4 & 0.00804935 & 0.0176695&   0.000716339\\
				\hline
				0.8& 0.80129&   0.932633&   0.181766\\
				\hline
				1.0 & 0.985614&   0.99713&   0.847873\\
				\hline
				1.2 & 0.99943&   0.999939&   0.989076\\
				\hline
				\multicolumn{4}{|c|}{ $l=1$, \,\,\,\,$n=0$,\,\,\,\, $\alpha=0.1$,\,\,\,\, $\Lambda=0.02$}\\
				\hline
				0& 0.0020804&   0.00297354&   0.000500328\\
				\hline
				0.4  & 0.0171343&   0.0377988&   0.00275507\\
				\hline
				0.8 & 0.817021&  0.937127&   0.307556\\
				\hline 
				1.0&  0.982834&   0.996419&   0.868466\\
				\hline
				1.2 &   0.999061& 0.999892&  0.987348\\
				\hline
				\multicolumn{4}{|c|}{$l=1$,\,\,\,\, $n=0$, \,\,\,\,$\alpha=0.15$, \,\,\,\,$\Lambda=0.02$ }\\
				\hline
				0&0.00186814&  0.0028915& 0.000138091 \\
				\hline
				0.4& 0.0157618 &0.0361107& 0.000928069 \\
				\hline
				0.8 &0.814444 &0.933686 &0.300518 \\
				\hline
				1.0& 0.983062 &0.996153 &0.902012\\
				\hline
				1.2 &0.999105 &0.999882&0.992818\\
				\hline
				\multicolumn{4}{|c|}{ $l=1$, \,\,\,\,$n=0$,\,\,\,\, $\alpha=0.15$,\,\,\,\, $\Lambda=0.08$}\\
				\hline
				0&0.00221037 &0.00275257 &0.000266319 \\
				\hline
				0.4 &0.0251606 &0.0502415 &0.00216865 \\
				\hline
				0.8 &0.906484  &0.968742 &0.52664\\
				\hline
				1.0 &0.993986  &0.998805 &0.951385\\
				\hline
				1.2 &0.999791 &0.999978  &0.997278\\
				\hline
				\multicolumn{4}{|c|}{ $l=1$, \,\,\,\,$n=0$,\,\,\,\, $\alpha=0.15$,\,\,\,\, $\Lambda=0.02$}\\
				\hline
				0&  0.00186903&  0.00268915&   0.000421534\\
				\hline
				0.4&  0.0157688&  0.0348879&  0.00237434\\
				\hline
				0.8& 0.814445&   0.936004&  0.290641\\
				\hline
				1.0& 0.983062&   0.99645&   0.865661\\
				\hline
				1.2& 0.999105&   0.999897&   0.987467\\
				\hline
				\multicolumn{4}{|c|}{ $l=2$, \,\,\,\,$n=0$,\,\,\,\, $\alpha=0.15$,\,\,\,\, $\Lambda=0.02$}\\
				\hline
				0&  0.0000821326&   0.000103032&   0.0000164771 \\
				\hline
				0.4&  0.000328439&   0.000454599&  0.000053879\\
				\hline
				0.8&  0.0205558&  0.0375335&   0.00187998\\
				\hline
				1.0&  0.301218&   0.468458&   0.0262424\\
				\hline
				1.2&  0.861667&  0.920698&   0.371926\\
				\hline
		\end{tabular}}
		\caption{The table shows the greybody factor of the scattered scalar, electromagnetic and Dirac wave  for different parameter values } 
		\label{tab:gf}
	\end{table}	
	
	\begin{table}
		\scalebox{0.8}
		{\renewcommand{\arraystretch}{1.5}
			\begin{tabular}{|p{2.7cm}|c|c|c|}
				\hline
				\multirow{2}{*}{\textit{Parameter changes}} & {\textit{Greybody factor $|T(\omega)|^2$} } & \multicolumn{2}{c|}{\textit{Quasinormal frequency}}\\
				\cline{3-4}
				&&\ \ \ \textit{Re} $(\omega)$ \ \ \  & \textit{Im} $(\omega)$\\
				\hline
				\multicolumn{4}{|c|}{\textit{Scalar case}}\\
				\hline
				\ \ \ \ \ \ \ \ $\alpha \uparrow$ & $\downarrow$ &  $\uparrow$ & $\uparrow$  \\
				\hline
				\ \ \ \ \ \ \ \ $\Lambda \uparrow$ & $\uparrow$ & $\downarrow$  &  $\uparrow$ \\
				\hline
				\ \ \ \ \ \ \ \ $l$\ \  $\uparrow$ & $\downarrow$ & $\uparrow$  & $\uparrow$ \\
				\hline
				\multicolumn{4}{|c|}{\textit{Electromagnetic case}}\\
				\hline
				\ \ \ \ \ \ \ \ $\alpha \uparrow$ & $\downarrow$ &  $\uparrow$ & $\uparrow$  \\
				\hline
				\ \ \ \ \ \ \ \ $\Lambda \uparrow$ & $\uparrow$ & $\downarrow$  &  $\uparrow$ \\
				\hline
				\ \ \ \ \ \ \ \ $l$\ \  $\uparrow$ & $\downarrow$ & $\uparrow$  & $\downarrow$ \\
				\hline
				\multicolumn{4}{|c|}{\textit{Dirac case}}\\
				\hline
				\ \ \ \ \ \ \ \ $\alpha \uparrow$ & $\downarrow$ &  $\uparrow$ & $\uparrow$  \\
				\hline
				\ \ \ \ \ \ \ \ $\Lambda \uparrow$ & $\uparrow$ & $\downarrow$  &  $\uparrow$ \\
				\hline
				\ \ \ \ \ \ \ \ $l$\ \  $\uparrow$ & $\downarrow$ & $\uparrow$  & $\downarrow$ \\
				\hline
		\end{tabular}}
		\caption{Qualitative changes in the greybody factor and the quasinormal frequency for all three cases i.e. scalar, electromagnetic and Dirac cases with increasing values of $\alpha$, $\Lambda$ and $l$. An increase/decrease in a particular quantity has been shown by an up/down arrow.}
		
		\label{tab:sum}
	\end{table}

	\section{Conclusion and future directions}
	Very recently, it has been shown \cite{Glavan:2019inb} that the EGB gravity theory can be reconstructed in a particular way where the Gauss-Bonnet coupling can be re-scaled as $\alpha/(D- 4)$. This theory in four space time dimension, the novel 4D EGB theory, defined as a $D\to 4$ limit at the level of equations of motion admits black hole solutions in asymptotically flat and (anti)-de Sitter spaces. The quasinormal modes of the scalar, gravitational and Fermionic fields for the asymptotically flat black holes in this background were already studied \cite{Konoplya:2020bxa, Churilova:2020aca}. Motivated by this, in this paper, we have extended the calculations to asymptotically de Sitter space time and evaluated the quasinormal modes of massless scalar, electromagnetic and Dirac field respectively. We summarise the results of our study in Table \ref{tab:sum}.
	
	We find that both the oscillation frequency and the damping time decreases with increasing values of the cosmological constant $\Lambda$. On the other hand, we observe that as the Gauss-Bonnet coupling $\alpha$ decrease and eventually crosses over to negative values, the real part of the frequency start decreasing whereas the imaginary part also starts to become more negative, implying that the damping increases. For positive increasing values of alpha the real part of the frequency increases. This remains the qualitative feature of all the three different types of perturbations that we have considered in this paper. From our results we can figure out the the stability of the scalar and electromagnetic perturbations can be confirmed from the positive definite potential, however, the Dirac case is a bit different.  The positive definiteness of the potential of any one of the potentials for any one of the chiralities does not help in asymptotically de Sitter black hole backgrounds because the potential for both chiralities will have negative gaps \cite{Konoplya:2020zso,Zinhailo:2019rwd}. Thus, one may require to perform a full time domain analysis in order to understand the stability feature of the space time under Dirac perturbation. The present study therefore can only give the qualitative nature of variations of the QN frequencies with the Gauss-Bonnet coupling and cosmological constant as far as fermionic perturbation is concerned. 
	
	Along with the quasinormal modes, we have also performed the calculation of the greybody factor for all three different types of perturbations. We have figured out that the general feature of the greybody factors for the three different types of perturbation fields is essentially similar. The greybody factor decreases with an increasing $l$ and an increase in the Gauss-Bonnet coupling constant $\alpha$. Finally, the greybody factors tend to increase with an increase in the cosmological constant. This behaviour could be easily explained by looking at Fig. \ref{fig:Veff}. The fraction of the wave transmitted upon scattering depends inversely on the height of the effective potential. With an increase in $\alpha$, the effective potential increases, for all three cases, and hence the transmission coefficient decreases. On the other hand with an increase in $\Lambda$, the effective potential decreases and hence the transmission coefficient increases. The dependence on $l$ could similarly be seen from Eq. (\ref{Vs}), Eq. (\ref{Vem}) and Eq. (\ref{Vd}). The effective potential for all three cases increases with an increase in $l$ and hence the transmission coefficient decreases.
	
	Novel four dimensional EGB gravity has created a lot of uproar ever since it was proposed. The importance of the theory lies in the fact that so far which was a higher dimensional theory (the Gauss-Bonnet term was only a topological term in four dimensions), can now be applied in the context of four dimensional space time in which we live in - this can open up many interesting windows in the study of alternative theories of gravity. Moreover, having a look at  the AdS branch will also be interesting in its own right. Calculations of the perturbations and the stability study of the novel 4D Gauss-Bonnet black hole in AdS background will be an important future extension of the present work. This may also be important to understand the AdS/CFT conjecture, since quasinormal modes describe the approach to equilibrium in the conformal field theory side. 
	
	{\it Note added: } On the day of the submission of the present manuscript, a paper appeared in arXiv \cite{Churilova:2020mif} which also deals with the same type of perturbations discussed here. While the paper \cite{Churilova:2020mif} gave the time domain analysis, which we did not present here, our work contains some more additional studies on greybody factor and eikonal limits.   
	

\end{document}